\documentclass[letter,12pt]{article}
\usepackage{jheppub}
\usepackage{epsfig,subfigure}
\usepackage{epstopdf}
\usepackage[utf8]{inputenc}
\usepackage{graphicx}
\usepackage{amssymb}
\usepackage{amsxtra}
\usepackage{amsmath}
\usepackage{booktabs,multirow,tabularx}
\usepackage{slashed}
\usepackage{float}
\usepackage{placeins}
\usepackage{rotating}
\usepackage{lscape}
\usepackage{color}
\usepackage{hyperref}
\usepackage{soul}

\newcommand{\bea}{\begin{eqnarray}}
\newcommand{\eea}{\end{eqnarray}}

\def\beq{\begin{equation}}
\def\eeq{\end{equation}}

\newcommand{\rom}[1]{\uppercase\expandafter{\romannumeral #1\relax}}

\DeclareUnicodeCharacter{2212}{-}

\preprint{PITT-PACC-2319}

\title{Probing the CP Structure of the Top Quark Yukawa at the Future Muon Collider
}

\author[a,b]{Morgan E. Cassidy,}
\author[a]{Zhongtian Dong,}
\author[a]{Kyoungchul Kong,}
\author[a]{Ian M. Lewis,}
\author[a,c]{Yanzhe Zhang,}
\author[a,d]{Ya-Juan Zheng}
\affiliation[a]{Department of Physics and Astronomy, University of Kansas, Lawrence, KS 66045, USA}
\affiliation[b]{Pittsburgh Particle Physics, Astrophysics, and Cosmology Center, Department of Physics and Astronomy, University of Pittsburgh, Pittsburgh, PA 15206, USA}
\affiliation[c]{Department of Astronomy,
University of Massachusetts Amherst,
Amherst, MA 01002, USA}
\affiliation[d]{Faculty of Education, Iwate University, Morioka, Iwate 020-8550, Japan}
\emailAdd{mec400@pitt.edu}
\emailAdd{cdong@ku.edu}
\emailAdd{kckong@ku.edu}
\emailAdd{ian.lewis@ku.edu}
\emailAdd{yanzhezhang@umass.edu}
\emailAdd{yjzheng@iwate-u.ac.jp}

\abstract{
We study the top-Higgs coupling with a CP violating phase $\xi$ at a future multi-TeV muon collider.  We focus on processes that are directly sensitive to the top quark Yukawa coupling: $t\bar{t}h$, $tbh\mu\nu$, and $t\bar{t}h\nu\bar{\nu}$ with $h\rightarrow b\bar{b}$ and semileptonic top decays. At different energies, different processes dominate the cross section, providing complementary information.  At and above an energy of $\mathcal{O}(10)$~TeV, vector boson fusion processes dominate.  As we show, in the Standard Model there is destructive interference in the vector boson fusion processes $t\bar{t}h\nu\bar{\nu}$ and $tbh\mu\nu$ between the top quark Yukawa and Higgs-gauge boson couplings. A CP-violating phase changes this interference, and the cross section measurement is very sensitive to the size of the CP-violating angle. Although we find that the cross sections are measured to $\mathcal{O}(50\%)$ statistical uncertainty at $1\sigma$, a 10 and 30 TeV muon collider can bound the CP-violating angle $|\xi|\lesssim9.0^\circ$ and $|\xi|\lesssim5.4^\circ$, respectively. However, cross section measurements are insensitive to the sign of the CP-violating angle.  To determine that the coupling is truly CP violating, observables sensitive to CP-violation must be measured.  We find in the $t\bar{t}h$ process the azimuthal angle between the $t+\bar{t}$ plane and the initial state muon+Higgs plane shows good discrimination for $\xi=\pm0.1\pi$. For the $tbh\mu\nu$ and $t\bar{t}h\nu\bar{\nu}$ processes, the operator proportional to $\left(\vec{p}_\mu\times\vec{p}_h\right)\cdot \vec{p}_t$ is sensitive to the sign of CP phase $\xi$.  From these observables, we construct asymmetry parameters that show good distinction between different values and signs of the CP violating angle.
  }

\begin{document}
\maketitle

%%%%%%%%%%%%%%%%%%%%%%%%%%%%%%%%%%%%%%%%%%%%%%%%%%%%%%
\section{Introduction}
\label{sec:intro}
%%%%%%%%%%%%%%%%%%%%%%%%%%%%%%%%%%%%%%%%%%%%%%%%%%%%%%
  One of the major goals of the Large Hadron Collider (LHC) and future collider programs is to measure the properties of the Higgs boson~\cite{Dawson:2022zbb,deBlas:2019rxi,Butler:2023eah,Narain:2022qud}.  Since the top quark is the heaviest particle in the Standard Model (SM) and has the strongest coupling to the Higgs, 
 processes involving the top quark Yukawa coupling provide us an ideal place to study the Higgs properties.  There has been much interest in directly probing the top Yukawa coupling at colliders.  This is typically done by measuring the Higgs produced in association with a top and anti-top ($t\bar{t}h$).  In the $h\rightarrow b\bar{b}$ channel, this was the first direct measurement of top-Higgs Yukawa interactions~\cite{CMS:2018uxb,ATLAS:2018mme}, constraining the Yukawa coupling magnitude to be within about $10\%$ of the SM prediction.   
 The high luminosity LHC (HL-LHC) is projected to measure the top quark Yukawa to within $3.4\%$~\cite{ATL-PHYS-PUB-2022-018,Dawson:2022zbb} of the SM value. 
When combined with the HL-LHC results, future electron positron colliders such as Higgs factories with energies of $240-500$~GeV~\cite{Bernardi:2022hny,CEPCPhysicsStudyGroup:2022uwl,deBlas:2019rxi} are expected to measure the top quark Yukawa with a precision of  $\sim3\%$~\cite{Dawson:2022zbb}.  Electron positron colliders with higher energies of $\mathcal{O}(1~{\rm TeV})$ and 100 TeV proton-proton machines will have a precision of $\sim1-2\%$~\cite{deBlas:2019rxi,Dawson:2022zbb}.

 Of particular interest is the charge conjugation and parity (CP) properties of the Higgs boson.       Higgs spin and CP properties are measured to be compatible with $J^{\rm PC}=0^{++}$ by the Higgs and vector bosons interactions ($hVV$) at the LHC~\cite{CMS:2019jdw, ATLAS:2018hxb}. However, the Higgs may still be an admixture of CP even and odd scalars, and different CP phases can appear in the Higgs-gauge boson couplings and the Higgs-fermion couplings.  In this paper we study the CP properties of the top quark Yukawa coupling at a future muon collider.  We parameterize the $ht\bar{t}$ interaction Lagrangian with a CP violating coupling via the relevant SMEFT~\cite{Buchmuller:1985jz,Grzadkowski:2010es} dimension-6 operator~\cite{Zhang:1994fb,Whisnant:1994fh,Whisnant:1997qu,Yang:1997iv,Barger:2023wbg} 
  \begin{eqnarray}
 \mathcal{L}_{htt,\text{SMEFT}} = -y_t \bar{Q}_L\widetilde{\Phi}t_R-c_t \Phi^\dagger \Phi \bar{Q}_L\widetilde{\Phi}t_R+{\rm h.c.}\label{eq:SMEFT} \, ,
 \end{eqnarray}
 where $y_t$ and $c_t$ are allowed to be complex, $Q_L=(t_L,b_L)^T$ is the left-handed quark doublet, and
 \begin{eqnarray}
 \Phi=\frac{1}{\sqrt{2}}\begin{pmatrix}0\\v+h\end{pmatrix}
 \end{eqnarray}
 is the Higgs doublet.  The Higgs boson is denoted by $h$ and $v=246$~GeV is the Higgs vacuum expectation value (vev).  A typical parameterization of the CP violating top Yukawa is through a CP violating phase $\xi$.  The SMEFT coupling constants can be rewritten via the identifications
 \begin{eqnarray}
 {\rm Re}(y_t)&=&-\frac{m_t}{\sqrt{2}\,v}\left(\kappa_{htt}\cos\xi-3\right)\nonumber \, , \\
 {\rm Re}(c_t)&=&\frac{\sqrt{2}m_t}{v^3}\left(\kappa_{htt}\cos\xi-1\right)\nonumber \, ,\\
 {\rm Im}(y_t)&=&-{\rm Im}(c_t)\frac{v^2}{2}=-\kappa_{htt}\frac{m_t}{\sqrt{2}\,v}\sin\xi \, .
 \end{eqnarray}
Using these, Eq.~(\ref{eq:SMEFT}) becomes
\footnote{The first line of Eq.~(\ref{eq:LCPangle}) is a typical parameterization of the CP violating top Yukawa in terms of the CP violating angle $\xi$.  However, as shown in Ref.~\cite{Barger:2023wbg}, the interaction $h-h-\bar{t}-t$ in the second line of Eq.~(\ref{eq:LCPangle}) contributes to $\mu^+\mu^-\to t\bar{t}h\nu\bar\nu$ and is necessary to obtain gauge invariant amplitudes in SMEFT with the dimension-6 operator.}
 \begin{eqnarray}
 \mathcal{L}_{htt,\text{SMEFT}}&=&-m_t\bar{t}{t}-g_{htt}h\bar{t}\left(\cos\xi+i\,\gamma_5\,\sin\xi\right)t\nonumber\\
 &&-\frac{3}{2}g_{htt}\frac{h^2}{v}\left(1+\frac{h}{3\,v}\right)\bar{t}\left(\cos\xi-\frac{1}{\kappa_{htt}}+i\,\gamma_5\,\sin\xi\right)t,\label{eq:LCPangle}
 \end{eqnarray}
with 
\begin{eqnarray}
g_{htt}\equiv(m_t/v)\kappa_{htt} \quad {\rm and} \quad -\pi\leq\xi\leq\pi,
\end{eqnarray}
where $m_t$ is the top quark mass.  
 When $\xi=0~(\pm\pi/2)$, Higgs couples with the top as a SM CP even scalar (CP odd pseudoscalar). The Yukawa coupling is CP violating in other cases.  The magnitude of the top Yukawa coupling $\kappa_{htt}$ can be bound by rate measurements of $t\bar{t}h$.  Due to the strong projected constraints at the HL-LHC we will assume $\kappa_{htt}=1$.

 Electron and neutron electric dipole moment  (EDM) bounds strongly constrain the CP odd component of Higgs-top quark Yukawa coupling to be $|\xi|\lesssim 0.08^\circ$~\cite{Brod:2022bww,Gritsan:2022php,Brod:2013cka,Bahl:2022yrs} at 95\% CL. However, these results rely on the Higgs coupling to first generation fermions being SM like.    Due to the very small couplings, it is extremely difficult to directly verify SM like couplings between the Higgs and electron and first generation quarks.  Allowing the Higgs couplings to first generation fermions to float drastically reduces the EDM bounds on the CP odd component of the top quark Yukawa coupling and the collider bounds are dominant~\cite{Brod:2013cka,Bahl:2022yrs}.  That is, without confirmation that the Higgs couplings to first generation fermions are indeed SM-like,  direct probes of the CP structure of the top quark Yukawa are needed.

 The single top with Higgs ($th$) production is very sensitive to the sign of top Yukawa coupling thanks to destructive interference between the $ht\bar{t}$ and $hWW$ couplings at leading order~\cite{Tait:2000sh,Maltoni:2001hu,Barger:2009ky,Demartin:2014fia,Barger:2018tqn,Barger:2019ccj}. The upper bound of the production rate of $th$ at the LHC is currently 12 times of the SM prediction~\cite{ATLAS:2020ior}.
The CP property has been studied in many different channels: $h \to \gamma\gamma$ \cite{ATLAS:2020ior}, $h \to b \bar b$ \cite{ATLAS:2023cbt}, multi-lepton final state \cite{CMS:2022dbt}.
 The combined analyses of $th$ and $t\bar{t}h$ from LHC Run {\rom1}~\cite{CMS:2015nrd} and Run {\rom 2} data~\cite{CMS:2018jeh,ATLAS:2020ior} suggest that the CP phase should be within the range of $|\xi|<43^\circ$ at 95$\%$ CL, and a pure CP odd coupling is excluded at 3.9$\sigma$.  There has been much interest in directly probing a CP violating top quark coupling at future colliders~\cite{Ellis:2013yxa,Boudjema:2015nda,Buckley:2015vsa,Demartin:2014fia,Gritsan:2016hjl,Mileo:2016mxg,AmorDosSantos:2017ayi,Azevedo:2017qiz,Li:2017dyz,Goncalves:2018agy,Ren:2019xhp,Bortolato:2020zcg,Cao:2020hhb,Martini:2021uey,Barman:2021yfh,Barman:2022pip,Gritsan:2022php,BhupalDev:2007ftb}.  
The CP violating angle can be bounded to be $|\xi|<25^\circ$ at the HL-LHC~\cite{Barman:2021yfh,Barman:2022pip} and $|\xi|<3^\circ$ at a 100 TeV pp collider~\cite{Goncalves:2021dcu,Barman:2022pip} at 95\% CL. 
We refer to Refs. \cite{Gritsan:2022php,Barman:2022pip,Ackerschott:2023nax,Heimel:2023mvw} on more details for potential improvement on the CP phase.

 In this work, we explore the CP violating $h\bar{t}t$ coupling via processes directly dependent on the top quark Yukawa at a future muon collider: $\mu^+\mu^-\rightarrow t\bar{t}h$, $\mu^+\mu^-\rightarrow t\bar{t}h\nu\bar{\nu}$ and $\mu^+\mu^-\rightarrow tbh\mu\nu${\footnote{Note that $tbhe\nu$ is not included in our study since it has only annihilation subprocess and therefore has a decreasing event rate with energy as well as sensitivity on $tth$ coupling.}}. We consider collider energies of $\sqrt{s}=1,\,3,\,10,$ and $30$~TeV.  A multi-TeV muon collider has been of interest for a long time~\cite{Palmer:1995jy,Palmer:2014nza,Boscolo:2018ytm,Delahaye:2019omf}, and there has been much interest recently~\cite{AlAli:2021let,Franceschini:2021aqd,Black:2022cth,MuonCollider:2022nsa,MuonCollider:2022xlm}. 
 Despite the technological challenges under investigation~\cite{Aime:2022flm,Accettura:2023ked,MuonCollider:2022glg,MuonCollider:2022nsa,MuonCollider:2022ded,Delahaye:2019omf}, the muon collider has the merit of both the high energy reach of a hadron collider and the cleanliness in background of the electron-positron collider.  Direct measurements of the top Yukawa in $t\bar{t}h$ production at a multi-TeV muon collider are projected to have an uncertainty of $35-50\%$~\cite{Forslund:2022xjq}.
 
 At multi-TeV energies and above, vector boson fusion (VBF) becomes the dominant production mode for many processes at muon colliders~\cite{Costantini:2020stv,AlAli:2021let,Han:2022edd,Aime:2022flm}.  In many SM processes, there is strong destructive interference in longitudinal VBF to guarantee unitarity.  Hence, small variations away from the predicted SM couplings can cause rates to grow very quickly with energy. Hence, rate measurements of VBF processes can provide strong bounds on new physics.   Indeed, by studying VBF production of top pairs $\mu^+\mu^-\rightarrow  t\bar{t}\nu\bar{\nu}$ with an off-shell Higgs, the top Yukawa can be indirectly measured to a precision of $6\%$~\cite{AlAli:2021let,Chen:2022yiu}.
 A recent study \cite{Liu:2023yrb} finds that a 10 TeV muon collider with an integrated luminosity of 10 ab$^{-1}$ could probe the top Yukawa coupling with a precision surpassing 1.5\%, more than one order of magnitude better than the precision from $t\bar t h\nu\bar{\nu}$ channel at muon colliders. 
 
We investigate the CP structure of the top Yukawa at a muon collider in a couple ways.  First, we study numerical and semi-analytic arguments about strong destructive interference in the SM calculation for the cross section of the VBF processes $\mu^+\mu^-\rightarrow t\bar{t}h\nu\bar{\nu},\,tbh\mu\nu$ and show that the CP violating angle can make this constructive interference.  However, the total cross section measurement is CP even with a CP violating angle.  Hence, to determine that a measured rate not in agreement with SM predictions is indeed due to a CP violating top Yukawa, we study observables that are genuinely sensitive to CP violation.  We then perform a full collider study for semi-leptonic top quark decays in $t\bar{t}h,\,t\bar{t}h\nu\bar{\nu}$ and hadronic top decay in $tbh\mu\nu$, with $h\rightarrow b\bar{b}$.  It should be noted that all these processes have the same final states and hence must be considered together.  Although we project a measurement of the total cross section with $\mathcal{O}(50\%)$ $1\sigma$ uncertainties at muon collider energies of $\gtrsim10$~TeV, we obtain very strong constraints on the CP violating angle from pure rate measurements.  

The paper is organized as follows. In Sec.~\ref{sec:prod} we study the event rates of $t\bar{t}h$, $tbh\mu\nu$ and $t\bar{t}h\nu\bar{\nu}$ in the SM and then explore the full $-\pi\leq\xi\leq\pi$ region of each process. We then study observables and construct asymmetries that can discriminate the magnitude and sign of $\xi$, i.e. observables that are genuinely sensitive to CP violation. In Sec.~\ref{sec:collider} we present the collider analysis with $h\rightarrow b\bar{b}$ and semileptonic decays.  The luminosity needed for discovery and exclusion of different $\xi$ are reported, as well as the expected 
exclusion at design luminosities.   Finally, in Sec.~\ref{sec:conc} we conclude.

%%%%%%%%%%%%%%%%%%%%%%%%%%%%%%%%%%%%%%%%%%%%%%%%%%%%%%
\section{Top-Higgs Production at a muon collider}
\label{sec:prod}
%%%%%%%%%%%%%%%%%%%%%%%%%%%%%%%%%%%%%%%%%%%%%%%%%%%%%%
\begin{figure}[tb]
\begin{center}
\subfigure[]{\includegraphics[width=0.25\textwidth,clip]{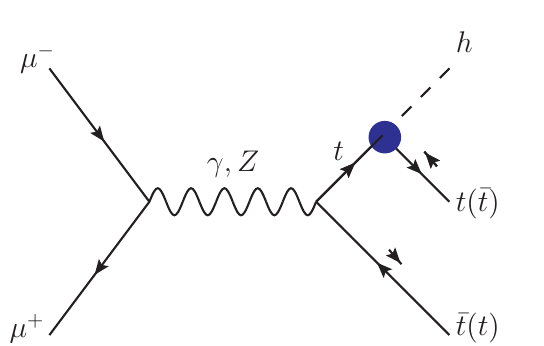}\label{fig:feyn_tth1}} 
\hspace*{1cm}
\subfigure[]{\includegraphics[width=0.25\textwidth,clip]{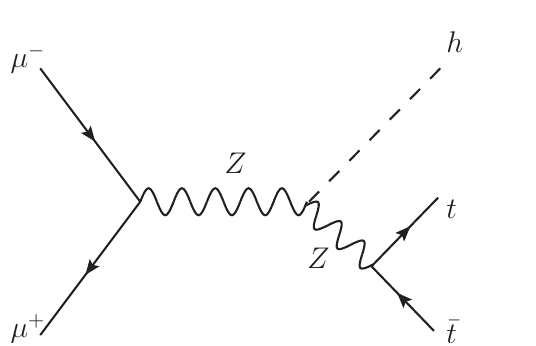}\label{fig:feyn_tth2}}
\end{center}
\caption{\label{fig:feyn_tth} Feynman diagrams for $\mu^+\mu^-\to t\bar{t}h$ production (a) with and (b) without a top quark Yukawa. Blue dots indicate the top Yukawa couplings.}
\end{figure}
\begin{figure}[tb]
\begin{center}
\subfigure[]{\includegraphics[width=0.25\textwidth,clip]{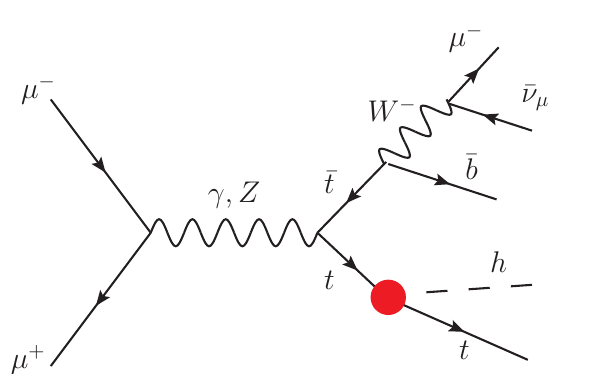}} \hspace*{0.5cm}
\label{fig:feyn_tbhmv_az}
\subfigure[]{\includegraphics[width=0.30\textwidth,clip]{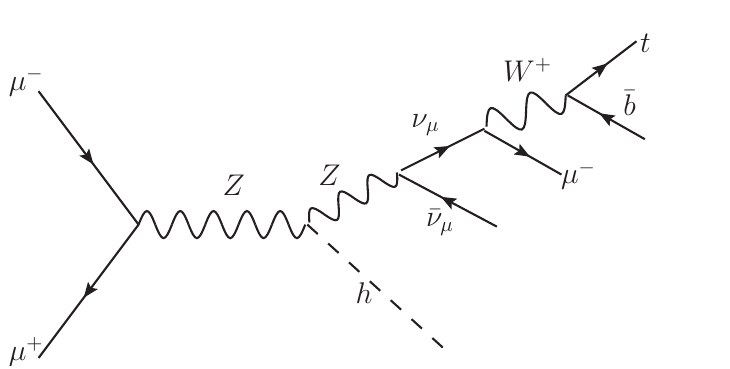}} \hspace*{0.5cm}
\label{fig:feyn_tbhmv_zh}
\subfigure[]{\includegraphics[width=0.25\textwidth,clip]{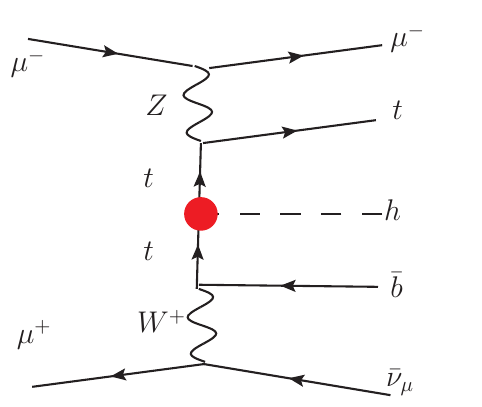}}  \\
\subfigure[]{\includegraphics[width=0.25\textwidth,clip]{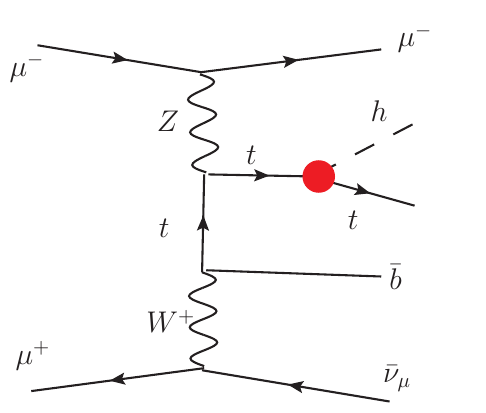}} \hspace*{0.5cm}
\subfigure[]{\includegraphics[width=0.25\textwidth,clip]{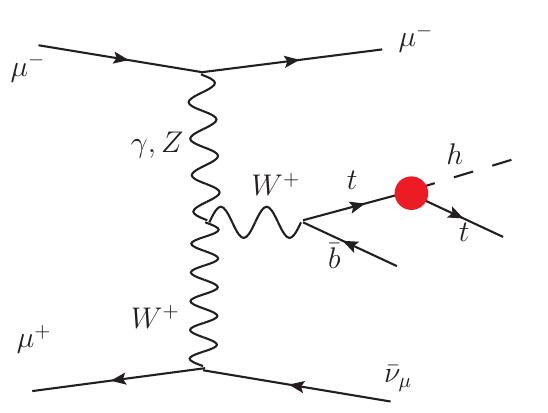}} \hspace*{0.5cm}
\subfigure[]{\includegraphics[width=0.25\textwidth,clip]{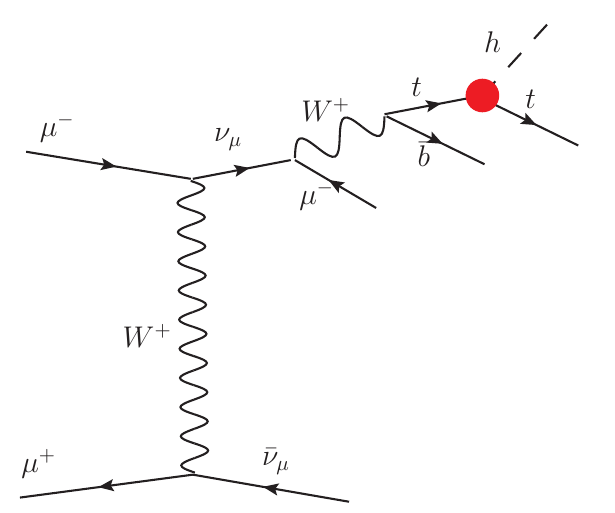}\label{fig:feyn_tbhmv_t_1int}}
\end{center}
\caption{\label{fig:feyn_tbhmv} Representative Feynman diagrams for $\mu^+\mu^-\to t\bar{b}h\mu^-\bar{\nu}$ production for $s$-channel type diagrams (a) with and (b) without top quark Yukawas; (c,d,e) VBF type diagrams with a top quark Yukawa; and (f) $t$-channel type diagram with a top quark Yukawa.  Red dots indicate the top Yukawa couplings.}
\end{figure}
\begin{figure}[t]
\begin{center}
\subfigure[]{\includegraphics[width=0.25\textwidth,clip]{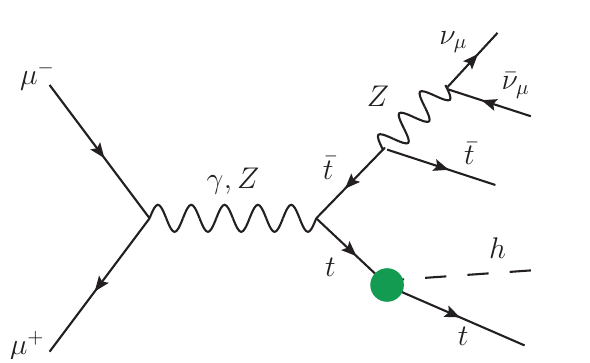}} \hspace*{0.5cm}
\subfigure[]{\includegraphics[width=0.30\textwidth,clip]{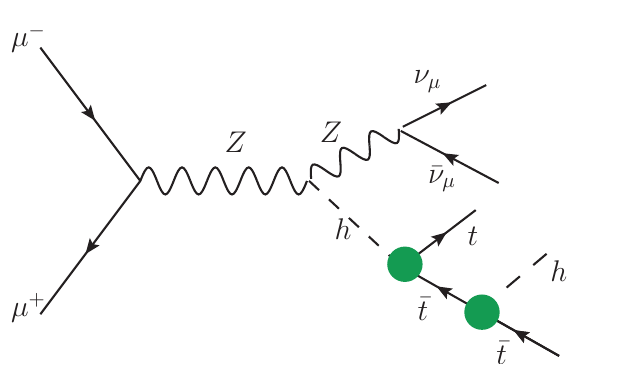}} \hspace*{0.5cm}
\subfigure[]{\includegraphics[width=0.25\textwidth,clip]{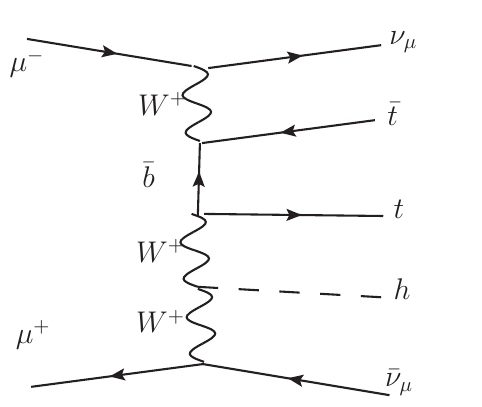}}\\
\subfigure[]{\includegraphics[width=0.25\textwidth,clip]{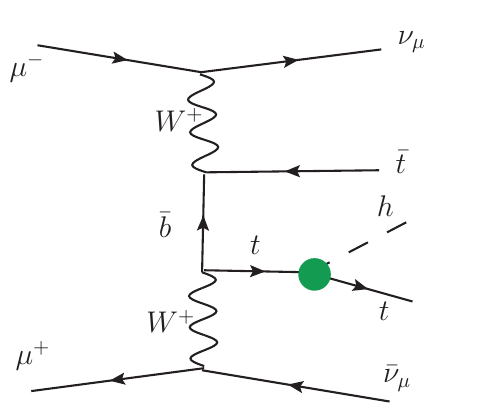}} \hspace*{0.5cm}
\subfigure[]{\includegraphics[width=0.25\textwidth,clip]{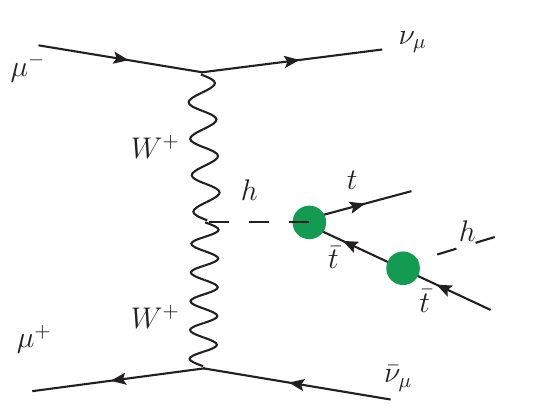}} \hspace*{0.5cm}
\subfigure[]{\includegraphics[width=0.25\textwidth,clip]{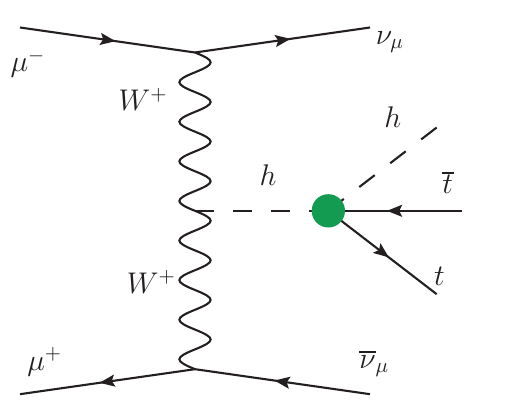}\label{fig:feyn_tthvv_t_tthh}}
\end{center}
\caption{ Representative Feynman diagrams for $\mu^+\mu^-\rightarrow t\bar{t}h\nu\bar{\nu}$ production for $s$-channel type diagrams with (a) one and (b) two top quark Yukawas; VBF type diagrams with (c) zero, (d) one, and (e) two top quark Yukawas and (f) one four point top-Higgs interaction from the dimension 6 operator.
Green dots indicate the top-Higgs couplings.
In addition to these diagrams, there are diagrams involving 4-point interactions with $W^+ W^- h h$ and $ZZhh$ and the Higgs trilinear coupling $hhh$ (not shown).
}
\label{fig:feyn}
\end{figure}

 In Figs.\,\ref{fig:feyn_tth},~\ref{fig:feyn_tbhmv}, and~\ref{fig:feyn} representative Feynman diagrams are shown for the production of 
 $\mu^-\mu^+\to t\bar{t}h, t\bar{b}h\mu^-\bar{\nu},$ and
$ t\bar{t}h\nu\bar{\nu}$, respectively.  The production of $t\bar{t}h$ is through s-channel photon and $Z$ diagrams, whereas production of $t\bar{t}h\nu\bar{\nu}$ and $t\bar{b}h\mu^-\bar{\nu}$ proceed through $s$-channel $\gamma/Z$, VBF, and $t$-channel type diagrams with vector boson exchange.  The $h-t-\bar{t}$ and $h-h-t-\bar{t}$ couplings are denoted by colored dots.  All three types of processes contain diagrams with no top Yukawa and a single top Yukawa.  The $t\bar{t}h\nu\bar{\nu}$ process also contains diagrams with two top Yukawa insertions, as illustrated in Figs.~\ref{fig:feyn}(b,e), {and with one 4-point  top-Higgs interaction, shown in Fig.~\ref{fig:feyn_tthvv_t_tthh}. }Although not shown, $t\bar{t}h\nu\bar{\nu}$ also contains diagrams sensitive to the quartic gauge boson-Higgs coupling $V-V-h-h$ or Higgs trilinear coupling $h-h-h$ where one Higgs splits into a $t\bar{t}$ pair.

\subsection{Pure CP even ($\xi=0$) and CP odd ($\xi=\pm\pi/2$) cases}
 
 \begin{figure}[tb]
\begin{center}
\includegraphics[width=0.65\textwidth,clip]{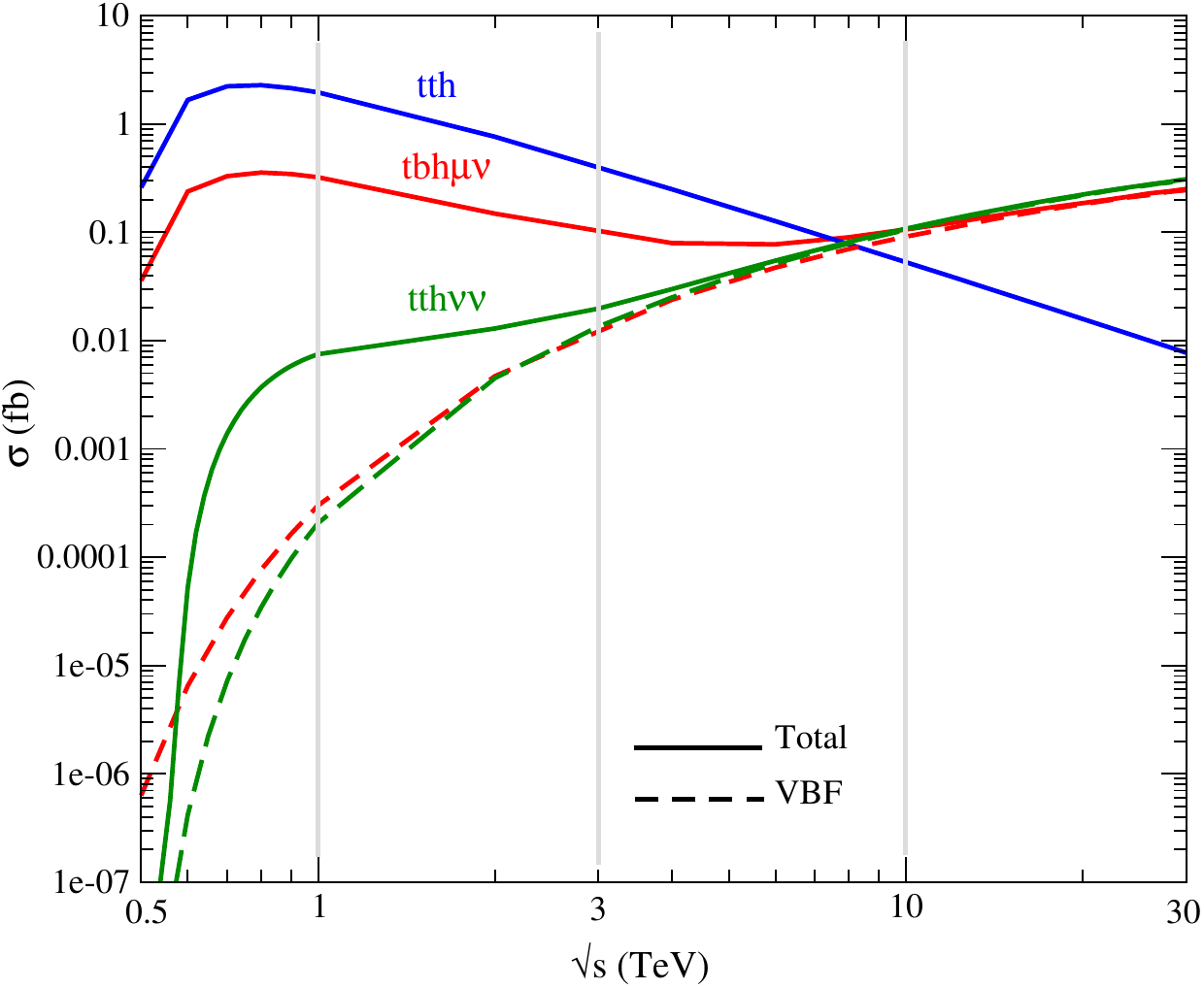}
\end{center}
\caption{\label{fig:xsSM} The SM ($\xi=0$) cross section for (blue) $\mu^-\mu^+\to t\bar{t}h$, (red) $tbh\mu\nu$ and (green) $t\bar{t}h\nu{\bar\nu}$ processes as a function of $\sqrt{s}$.  Solid lines are the total cross sections and dashed lines are the VBF contributions.}
\label{fig:smxs}
\end{figure}

 We show in Fig.\,\ref{fig:smxs} the SM ($\xi=0$) cross section as a function of center of momentum energy, $\sqrt{s}$, for
  \begin{eqnarray}
  \rm{(blue)}~\mu^+\mu^-&\to& t\bar{t}h, 
  \nonumber
  \\
   \rm{(green)}~\mu^+\mu^-&\to& t\bar{t}h\nu{\bar\nu}\equiv t\bar{t}h\nu_\ell{\bar\nu}_\ell\quad(\ell=e,~ \mu~{\rm and}~\tau),\quad{\rm and}
   \nonumber
   \\
  \rm{(red)}~\mu^+\mu^-&\to &  tbh\mu\nu\equiv t\bar{b}h\mu^-\nu_\mu+\bar{t}{b}h\mu^+\bar{\nu}_\mu,
  \end{eqnarray}
  with $\sqrt{s}$ from 500 GeV to 30 TeV.  The conjugate processes $t\bar{b}h\mu^-\bar{\nu}$ and $\bar{t}bh\mu^+\nu$ 
  are combined.  The cross sections are generated with {\tt MadGraph5\_aMC@NLO}\cite{Alwall:2014hca}.  
  The total cross section is shown as solid lines and the VBF subprocess contributions are shown separately in dashed lines for $tbh\mu\nu$ and $t\bar{t}h\nu\bar{\nu}$.  To insure a gauge invariant subset of diagrams, the VBF subprocesses are isolated by replacing the initial state $\mu^+$ with an $e^+$ as adopted in Ref.\,\cite{Costantini:2020stv}.    Even though the attraction of a muon collider is to have a multi-TeV lepton collider, we consider $\sqrt{s}\gtrsim0.5$~TeV to gauge our understanding and reach at lower energy.  We terminate the plot at $\sqrt{s}=30$~TeV since for larger energies the electroweak (EW) Sudakov logarithms in the VBF diagrams with massive vector bosons become large and the effective vector boson approximation is needed for a reliable calculation of rates~\cite{Dawson:1984gx,Ruiz:2021tdt,Costantini:2020stv,Han:2020uid}. For $tbh\mu\nu$ production, the $t$-channel type diagrams 
  can also be realized by a photon exchange.  The massless photon mediator in this case would cause a singularity when a final state muon is collinear with an initial state muon.  A proper treatment also involves the effective vector boson approximation~\cite{vonWeizsacker:1934nji,Williams:1935dka,Dawson:1984gx,Han:2020uid,Ruiz:2021tdt}. We impose a cut on the transverse momentum of the outgoing muon $p_T^{\mu}>10$~GeV
  to regulate the singularity and avoid the complexity of the effective vector boson approximation.
  
  The $t\bar{t}h$ process has only $s$-channel contributions.  Therefore, its cross section decreases as the energy of the collider increases past the mass threshold, from about $2.0$~fb at $\sqrt{s}=1$~TeV to $8\times10^{-3}$~fb at $\sqrt{s}=30$ TeV.  As mentioned previously, the $t\bar{t}h\nu\bar{\nu}$ and $tbh\mu\nu$ have both $s$-channel and VBF contributions.  When the $s$-channel contributions are dominant, the cross sections decrease similarly to $t\bar{t}h$.  Once the VBF contributions are dominant, there is an EW Sudakov $\log$ enhancement that causes the cross sections to increase at high energy. For $t\bar{t}h\nu\bar{\nu}$, the VBF contributions become dominant at $\sqrt{s}\gtrsim3$ TeV, while it is from $\sqrt{s}\gtrsim5 $ TeV that $tbh\mu\nu$ receives contribution mainly from VBF subprocesses.

\begin{table}[t]
\begin{center}
\begin{tabular}{|c|cccc|cccc|}\hline
&\multicolumn{4}{c|}{CP even} &\multicolumn{4}{c|}{CP odd}\\
\hline
$\sqrt{s} ~({\rm TeV})$&1&3&10&30&1&3&10&30\\
$\int{\cal L}dt~ (fb^{-1})$&$10^2$&$10^3$&$10^4$&$10^4$&$10^2$&$10^3$&$10^4$&$10^4$\\
\hline
$N(t\bar{t}h)$&210&420&$550$&$80$&51&230&400&63\\
%\hline
$N(tbh\mu\nu)$&43&100&1,300&3,100&11&120&5,400&$1.7\cdot10^4$\\
$N(t\bar{t}h\nu\bar{\nu})$&1&19&980&3,100&$<1$&180&$2.9\cdot10^4$&$2.4\cdot10^5$\\
\hline
\end{tabular}
\end{center}
%}
\caption{\label{tab:lumi} Target 5-year integrated luminosities and expected number of events for $t\bar{t}h$, $t{b}h\mu{\nu}$ and $t\bar{t}h\nu\bar{\nu}$ in the pure CP even ($\xi=0$) and pure CP odd ($\xi=\pm\pi/2$) state. }
\end{table}

  In the following discussion, we will choose $\sqrt{s}=$ 1, 3, 10 and 30 TeV as the benchmark energies.  Table~\ref{tab:lumi} shows the benchmark luminosities and expected number of events at the benchmark energies for pure CP even ($\xi=0$) and CP odd ($\xi=\pm\pi/2$) cases. These rates are estimated using 100,000 events generated in \texttt{MadGraph5\_aMC@NLO}~\cite{Alwall:2014hca}.  The signal model is implemented using \texttt{FeynRules}~\cite{Christensen:2008py,Alloul:2013bka}.  For energies at and below 10 TeV, the benchmark luminosities are set by the minimal  instantaneous luminosity needed to generate 10,000 events for a process with 1 fb cross section  within 5 years at $\sqrt{s}=10$~TeV~\cite{Delahaye:2019omf}:
\begin{eqnarray}
{\cal L}\gtrsim\frac{{\rm 5~years}}{{\rm time}}\left(\frac{\sqrt{s}}{\rm 10~TeV}\right)^22\times10^{35}{\rm cm^{-2}{\rm s^{-1}}}.
\end{eqnarray}
This is the number of events needed for percent level statistical uncertainty.
For collider energies above $10$~TeV, we consider the luminosity benchmarks based on the conservative designs of Table 1 of Ref. \cite{AlAli:2021let}.

 For $tbh\mu\nu$ and $t\bar{t}h\nu\bar{\nu}$, at $\sqrt{s}=10$ and $30$~TeV we can expect $10^3$ events for the CP even state and $10^3-10^5$ events for the CP odd state.  What is particularly striking is at 10 and 30 TeV, the pure CP odd case has significantly more events than the pure CP even case in the $tbh\mu\nu$ and $t\bar{t}h\nu\bar{\nu}$  channels.  We will discuss this phenomena in depth below.

\subsection{CP admixtures}

\begin{figure}[tb]
\begin{center}
\subfigure[]{\includegraphics[width=0.47\textwidth,clip]{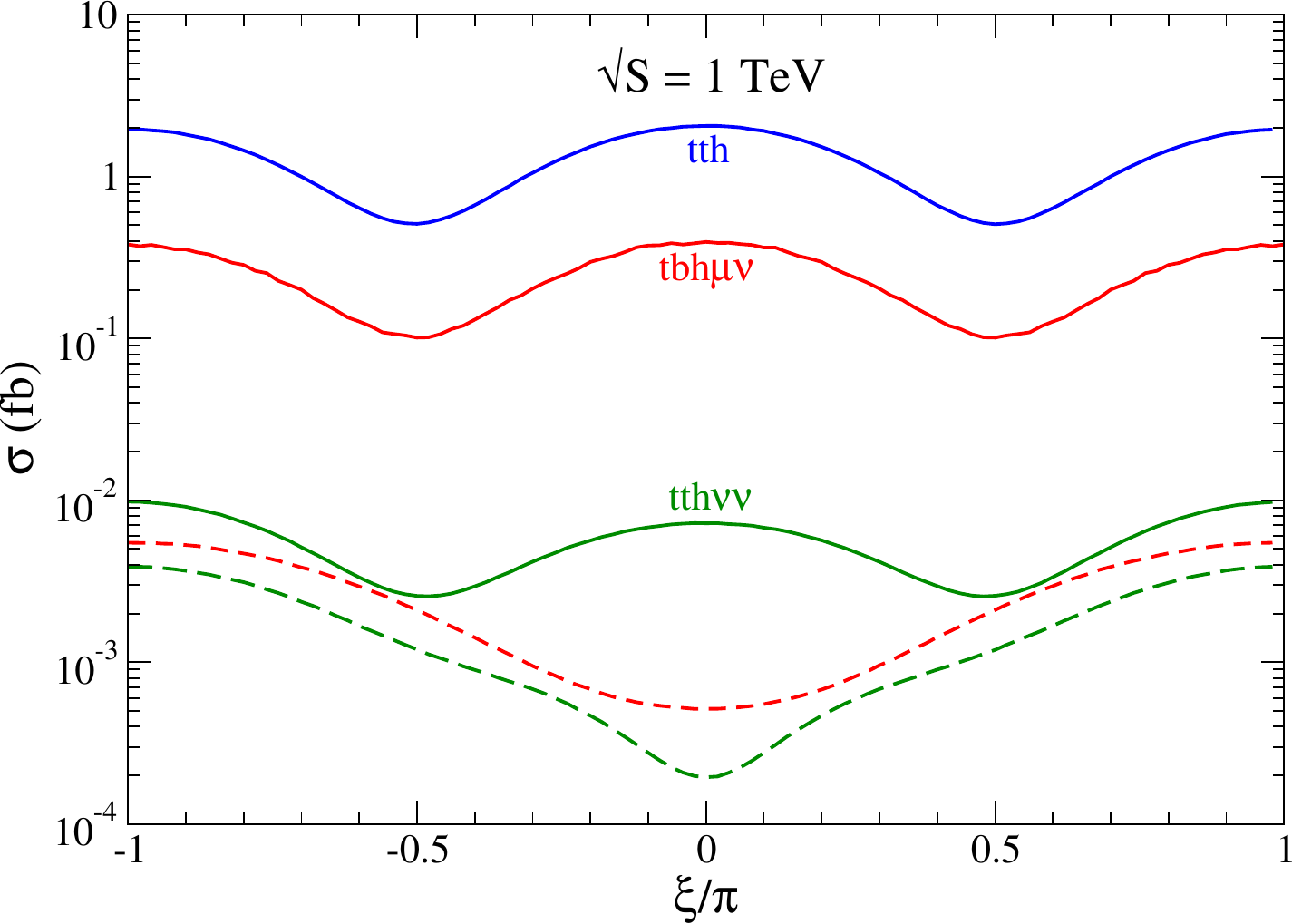}} \hspace*{0.1cm}
\subfigure[]{\includegraphics[width=0.47\textwidth,clip]{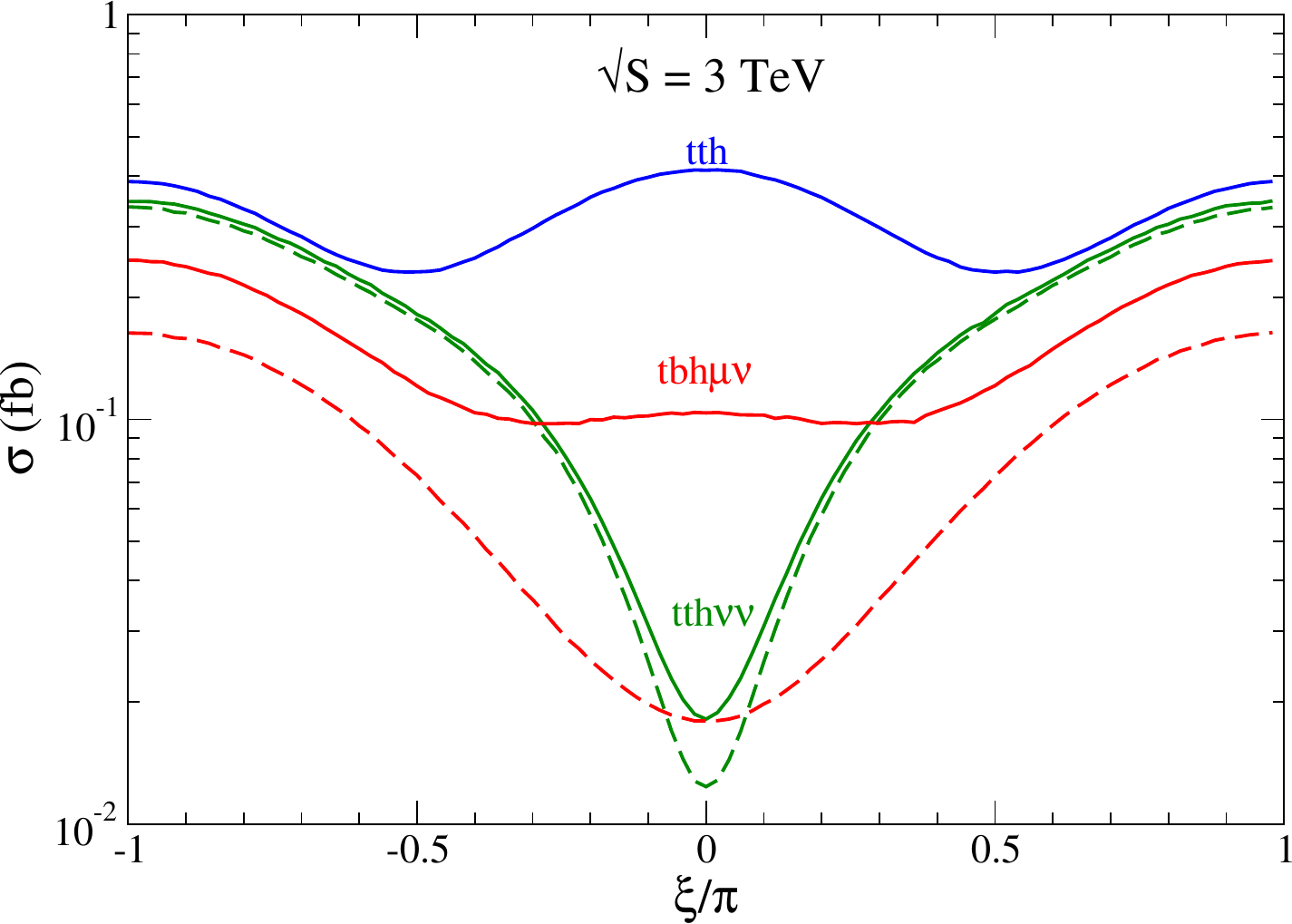}}
\subfigure[]{\includegraphics[width=0.47\textwidth,clip]{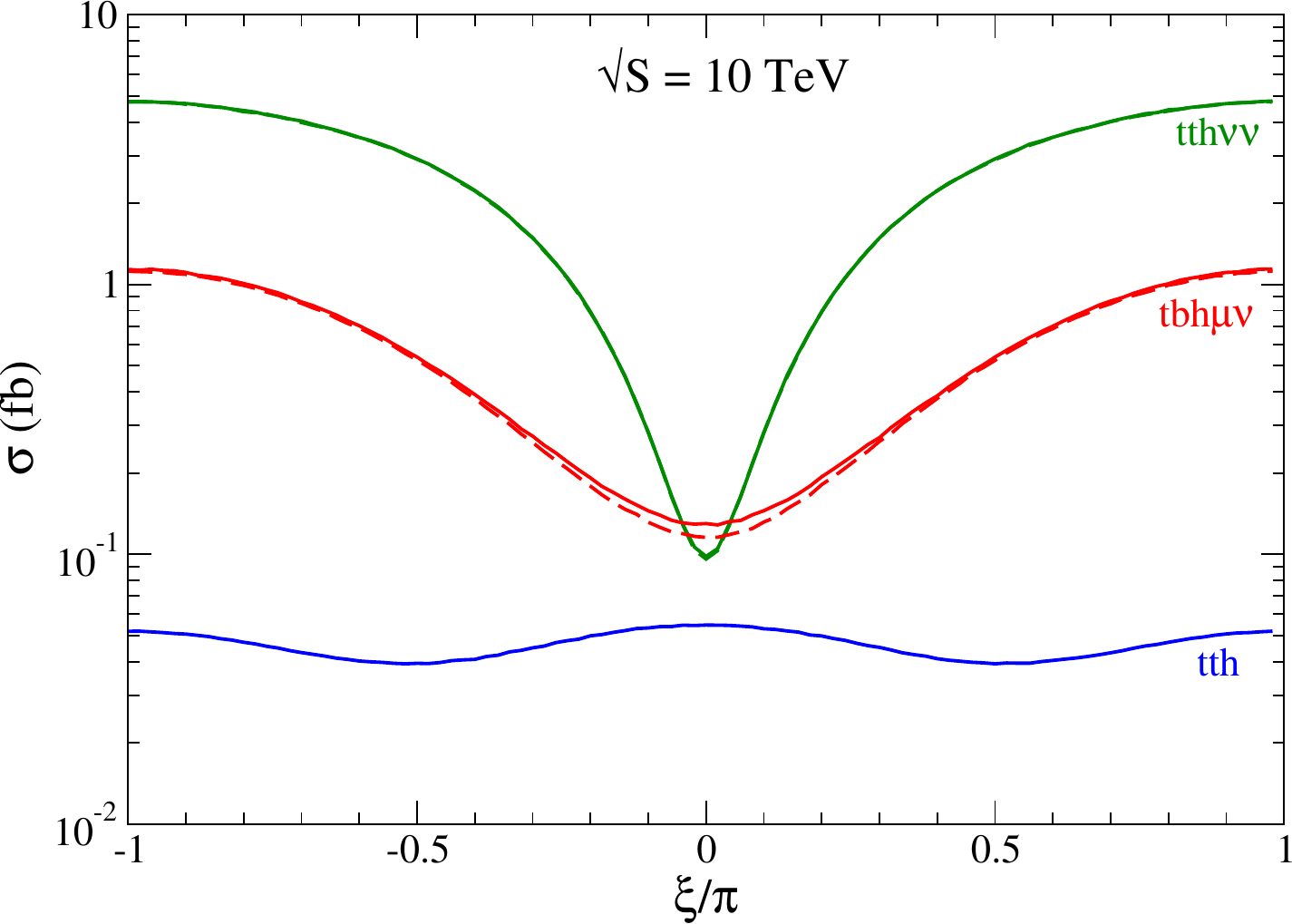}} \hspace*{0.1cm}
\subfigure[]{\includegraphics[width=0.47\textwidth,clip]{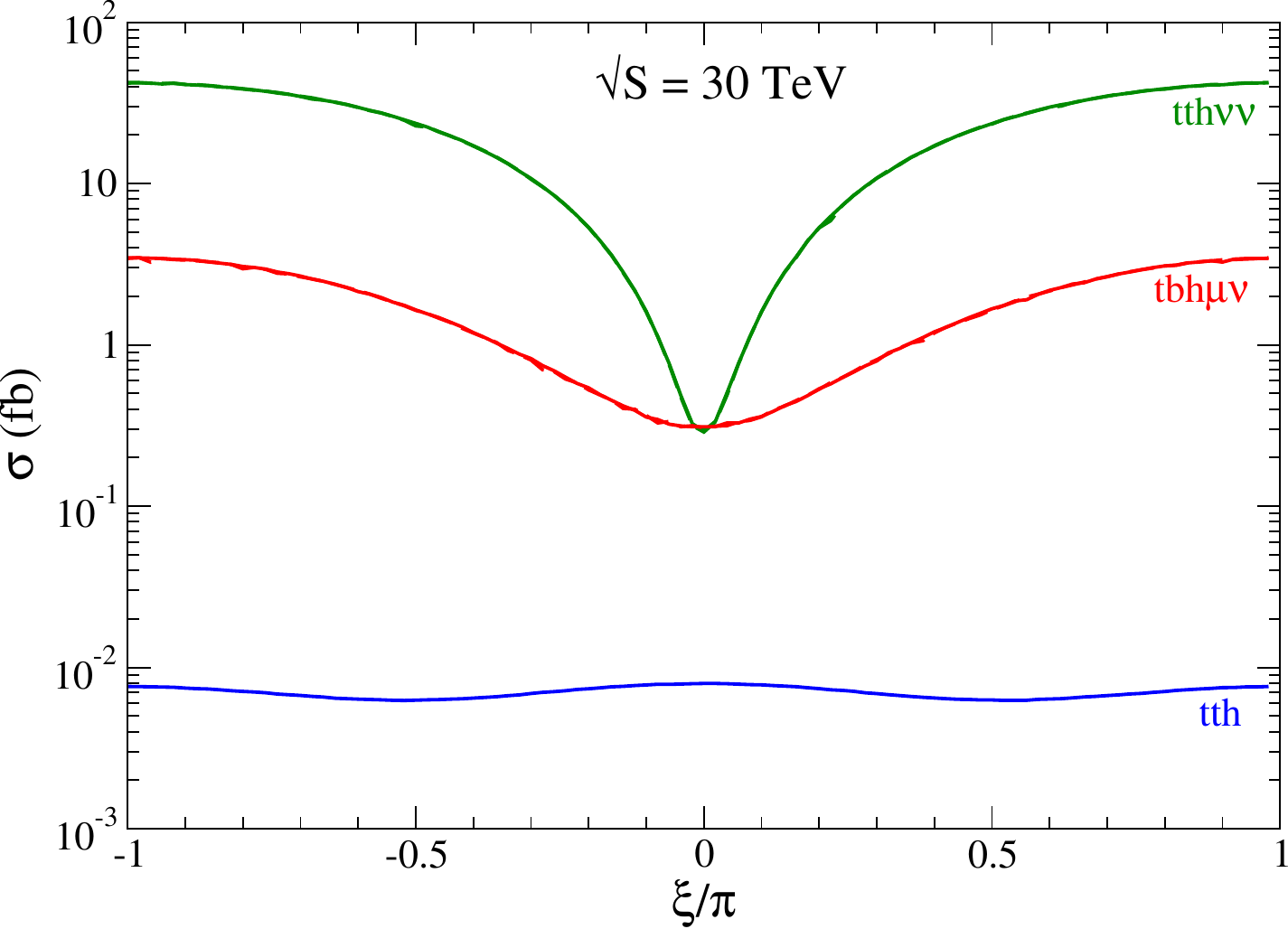}}
\end{center}
\vspace*{-0.5cm}
\caption{\label{fig:xsCPV} The cross section as a function of the CP violating phase $\xi$ for (blue) $\mu^+\mu^-\to t\bar{t}h$, (green) $t\bar{t}h\nu{\bar\nu}$, and (red) $tbh\mu\nu$ at (a) 1 TeV, (b) 3 TeV, (c) 10 TeV, and (d) 30 TeV. A cut of $p_T^{\mu}>10$ GeV is applied for $tbh\mu\nu$.  Solid lines indicate total cross section while dashed lines are the VBF contribution.}
\end{figure}

Figure~\ref{fig:xsCPV} shows the cross sections for $t\bar{t}h$, $t\bar{t}h\nu\bar{\nu}$ and $tbh\mu\nu$ as functions of the CP violating phase at (a) 1 TeV,  (b) 3 TeV, (c) 10 TeV, and (d) 30 TeV. Green and red dashed lines represent the VBF contributions for $t\bar{t}h\nu\bar{\nu}$ and $tbh\mu\nu$.   The $s$-channel and VBF style diagrams have different dependencies on $\xi$.  As can be seen, for $s$-channel processes the SM ($\xi=0$) has the maximum cross section, while for processes dominated by VBF {the wrong sign Yukawa  $\xi=\pm \pi$ cases} have the maximum cross section.  Also, the periodicity of the cross section dependence on $\xi$ is different for $s$-channel and VBF contributions.  Finally, as the center of momentum energy of the collider increases the cross section dependence on $\xi$ becomes very strong for the VBF dominant  processes.  

To understand the cross section dependence on $\xi$, we take an semi-analytical approach.  First we discuss the $t\bar{t}h$ production.  Since the total cross section is CP even, it can only depend on $\cos^2\xi$, $\sin^2\xi$, $\cos\xi$, and a piece independent of $\xi$.  The terms proportional to $\cos^2\xi$ and $\sin^2\xi$ originate from the modulus square of diagrams where the Higgs is radiated off a top quark, such as Fig.~\ref{fig:feyn_tth1}. The terms independent of $\xi$ are from the square of diagrams where the Higgs is radiated off an internal $Z$-boson\footnote{Due to its very small value, we assume the muon Yukawa is negligible.}, such as Fig.~\ref{fig:feyn_tth2}. Finally, the terms proportional to $\cos\xi$ are due to interference between diagrams where the Higgs is radiated off of a top quark or $Z$-boson. Hence, for $t\bar{t}h$ production we use the general parameterization,
 \begin{eqnarray}
 \sigma_{t\bar{t}h}(\xi)=C^2_{tth}\cos^2\xi+C^1_{tth}\cos\xi+C^0_{tth}.
 \label{eq:tthxs}
 \end{eqnarray}
 The numerical results of the coefficients $C_{tth}^i\,(i=0,1,2)$ at $\sqrt{s}=1$, 3, 10 and 30 TeV are summarized in Table \,\ref{tab:C43210}. All results in Tab.~\ref{tab:C43210} are found by scanning over $\xi$ in~\texttt{MadGraph5\_aMC@NLO}~\cite{Alwall:2014hca} with 50,000 events at each point. The stability fit is checked by also fitting with additional higher powers to $\cos\xi$ and insuring the fit does not change. The dominance of the $\cos^2\xi$ and constant terms explain the overall shape of the $t\bar{t}h$ cross section dependence on $\xi$, which has an approximate $\pi/2$ periodicity.  As the collider energy increases the constant term $C^0_{tth}$ becomes the largest coefficient decreasing the cross section dependence on $\xi$. The non-zero value of the $\cos\xi$ coefficient explains the small asymmetry between $\xi=0$ and $\xi=\pm\pi$, which slightly breaks the $\pi/2$ periodicity and reveals the weak interference between the diagrams with $htt$ and $hZZ$ couplings in Fig.\,\ref{fig:feyn_tth}(a) and (b), respectively

\begin{table}
\begin{center}
\resizebox{\linewidth}{!}{%
\begin{tabular}{|c|c||ccccc|}\hline
Process&$\sqrt{s}$ (TeV) &$C^0$ (fb) & $C^1$ (fb) & $C^2$ (fb) & $C^3$ (fb) & $C^4$ (fb)\\\hline\hline
\multirow{4}{*}{$t\bar{t}h$} & $1$ & $0.511$ & $0.0465$ & $1.523$ & - & -\\
&$3$ & $0.233$ & $0.0134$ & $0.173$ & - & -\\
&$10$ & $0.0397$ & $1.48\cdot 10^{-3}$ & $0.0142$ &- &- \\
& $30$ & $6.33\cdot10^{-3}$&$1.69\cdot10^{-4}$ & $1.53\cdot10^{-3}$ & - & -\\\hline\hline
\multirow{4}{*}{$tbh\mu\nu$} & 
$1$ & $0.108$ & $8.29\cdot10^{-3}$  & $0.312$ & - & -\\
& $3$& $0.122$ & $-0.0717$  & $0.0537$  & - & -\\
&$10$ & $0.537$ & $-0.503$ & $0.0967$ &- &- \\
& $30$ &$1.66$ & $-1.57$ & $0.224$ & - & -\\\hline\hline
\multirow{4}{*}{$t\bar{t}h\nu\bar{\nu}$} & 
$1$ & $2.57\cdot10^{-3}$ & $-5.83\cdot10^{-4}$  & $6.12\cdot10^{-3}$  & $-6.88\cdot10^{-4}$ & $-1.70\cdot10^{-4}$\\
& $3$&$0.184$  &$-0.119$  &$4.25\cdot10^{-3}$  & $-0.0468$ & $-4.50\cdot10^{-3}$\\
&$10$ & $2.92$ & $-2.06$  & $-0.468$ & $-0.280$ & $-0.0149$ \\
& $30$ & $23.9$ &$-20.4$ &$-2.52$ & $-0.712$ & $0.0446$\\\hline\hline
\end{tabular}
}
\end{center}
\caption{\label{tab:C43210}Coefficients for $t\bar{t}h,tbh\mu\nu$ and $t\bar{t}h\nu\bar{\nu}$ cross section parameterizations in Eqs.~(\ref{eq:tthxs},~\ref{eq:xsect_tbhmn},~\ref{eq:tthvvxs}).}
\end{table}

 For $tbh\mu\nu$ production, there are still at most single insertions of the top quark Yukawa.  Hence, we can use a similar parameterization as in the $t\bar{t}h$ production cross section:
 \begin{eqnarray}
 \sigma_{tbh\mu\nu}(\xi)=C_{tbh\mu\nu}^2\cos^2\xi+C_{tbh\mu\nu}^1\cos\xi+C_{tbh\mu\nu}^0.\label{eq:xsect_tbhmn}
 \end{eqnarray}
 Unlike the $t\bar{t}h$, there are also diagrams where the Higgs is radiated off a $b$-quark or a $W$-boson.  Together with radiation off the $Z$-boson, these interactions contribute to the interference term $C_{tbh\mu\nu}^1$ and the constant term $C_{tbh\mu\nu}^0$.  
 The numerical results are also shown in Table \ref{tab:C43210}.  At $\sqrt{s}=1$~TeV, the coefficients $C^2_{tbh\mu\nu}$ and $C^0_{tbh\mu\nu}$ are largest.  At this energy the $s$-channel diagrams dominate, hence the $\xi$ dependence is similar to $t\bar{t}h$.  At higher energies, the VBF diagrams dominate and the interference term $C^1_{tbh\mu\nu}$ becomes more important.  Indeed, it can be seen in the SM case ($\xi=0$) there is strong destructive interference between the top Yukawa and other Higgs couplings.  Due to the smallness of the bottom quark mass, the major contribution to this interference is between the top and massive vector boson couplings with the Higgs. When we move away from the SM limit to {the wrong sign Yukawa} case ($\xi=\pm \pi)$, the strong destructive interference becomes constructive interference. The cross sections at $\sqrt{s}=30$~TeV are
 \begin{eqnarray}
 \sigma_{tbh\mu\nu}(\xi=0)=0.31~{\rm fb}
\quad\quad{\rm and} \quad\quad
 \sigma_{tbh\mu\nu}(\xi=\pm\pi)=3.5~{\rm fb}.
 \end{eqnarray}
 Cross sections at $\xi=\pm\pi$ are an order of magnitude larger than those at $\xi=0$.   
 This explains why at high energy the $tbh\mu\nu$ rate depends strongly on $\xi$. It should be noted, similar interference effects between the top and massive vector boson couplings to the Higgs are seen in single $t$+$h$ production at the LHC~\cite{Tait:2000sh,Maltoni:2001hu,Barger:2009ky,Demartin:2014fia,Barger:2018tqn,Barger:2019ccj}.

 The cross section dependence on $\xi$ for $t\bar{t}h\nu\bar{\nu}$ is different than the dependence of the $t\bar{t}h$ and $tbh\mu\nu$ cross sections.  There are now diagrams with intermediate Higgs bosons that have double insertions of the top Yukawa, as seen in Figs.~\ref{fig:feyn}(b,e).  Additionally, some of these diagrams depend on the coupling between the Higgs and massive vector bosons, potentially complicating the interference story.  Keeping in mind that the total cross section is CP even, the double insertion of the top Yukawa introduces new dependence on $\xi$.  The modulus square of the diagrams with two top Yukawas contribute  $\cos^4\xi,\,\sin^4\xi,\,\cos^2 \xi \sin^2\xi$.  
 {There are also diagrams with a single 4-point $h-h-t-\bar{t}$ coupling from the dimension-6 operator as shown in Fig.~\ref{fig:feyn_tthvv_t_tthh}.  However, the dependence of the four point top-Higgs coupling on $\xi$ is similar to that as the top Yukawa, i.e.\,$\propto\cos\xi+i\gamma_5\sin\xi-1$, as can be seen in Eq.~(\ref{eq:LCPangle}).  Hence, this four point coupling does not introduce new dependencies on $\xi$ in the cross section.} 
 The interference between diagrams with two top Yukawas and other processes then contribute to $\cos^3\xi,\,\cos\xi\sin^2\xi,\,\cos^2\xi,\,$ and $\sin^2\xi$.  We parameterize the $\xi$ dependence as
  \begin{eqnarray}
\sigma(t\bar{t}h\nu\bar{\nu})=C^4_{tth\nu\nu}\cos^4\xi+C^3_{tth\nu\nu}\cos^3\xi+C^2_{tth\nu\nu}\cos^2\xi+C^1_{tth\nu\nu}\cos\xi+C^0_{t{t}h\nu{\nu}}.
 \label{eq:tthvvxs}
 \end{eqnarray}
 It should be noted, that unlike $t\bar th$ and $tbh\mu\nu$, the $t\bar{t}h\nu\bar{\nu}$ process also depends on the Higgs trilinear coupling $h-h-h$ and the quartic coupling between the Higgs boson and massive vector boson $h-h-V-V$.

 The numerical values of the coefficients $C^i_{tth\nu\nu}$ are given in Table~\ref{tab:C43210}. 
 At all energies under consideration, there is destructive interference.  The destructive interference becomes very strong at high energies where the numerical values of $C^1_{tth\nu\nu}$ and $C^0_{tth\nu\nu}$ are the largest.  At $\sqrt{s}=30$ TeV, the SM cross section is 
 \begin{eqnarray}
 \sigma_{tth\nu\nu}(\xi=0)=0.31~{\rm fb}.
 \end{eqnarray}
 The {wrong sign Yukawa} cross section is
 \begin{eqnarray}
 \sigma_{tth\nu\nu}(\xi=\pm \pi)=43~{\rm fb}.
 \end{eqnarray}
 The $\xi=\pm \pi$ cross section is two orders of magnitude larger than the $\xi=0$ cross section. 
 This $\xi$ dependence explains why at 10 and 30~TeV, the $t\bar{t}h\nu\bar{\nu}$ cross section has the strongest dependence on $\xi$  (see Fig.~\ref{fig:xsCPV}).  It is clear that a pure rate measurement of top+Higgs production at a high energy muon collider is very sensitive to non-zero $\xi$.

It is worth noting that at hadron colliders a direct Higgs-top coupling measurement is from $gg(q\bar{q})\to t\bar{t}h$ production and all the contributed diagrams depend on $\xi$ at leading order.  Hence, the total $t\bar{t}h$ cross section at hadron colliders is symmetric between $\xi=0$ versus $\xi=\pm\pi$.  {As discussed earlier, the rate measurement of single top+Higgs production at hadron colliders is sensitive to the difference between the SM ($\xi=0$) and wrong sign Yukawa ($\xi=\pm\pi$) cases~\cite{Tait:2000sh,Maltoni:2001hu,Barger:2009ky,Demartin:2014fia,Barger:2018tqn,Barger:2019ccj}.  However, this rate is quite small making it difficult to directly measure small values $\xi$ in this process in the current LHC run.}  This is unlike the case at lepton colliders with $\sqrt{s}\gtrsim \mathcal{O}(10~{\rm TeV})$, where we can easily disentangle the $\xi=0$ and $\xi\neq0$ just from rate measurements.  We will quantify this in Sec.~\ref{sec:collider}.
 
\subsection{CP violating observables }

 \begin{figure}[tb]
\begin{center}
\subfigure[]{\includegraphics[width=0.47\textwidth,clip]{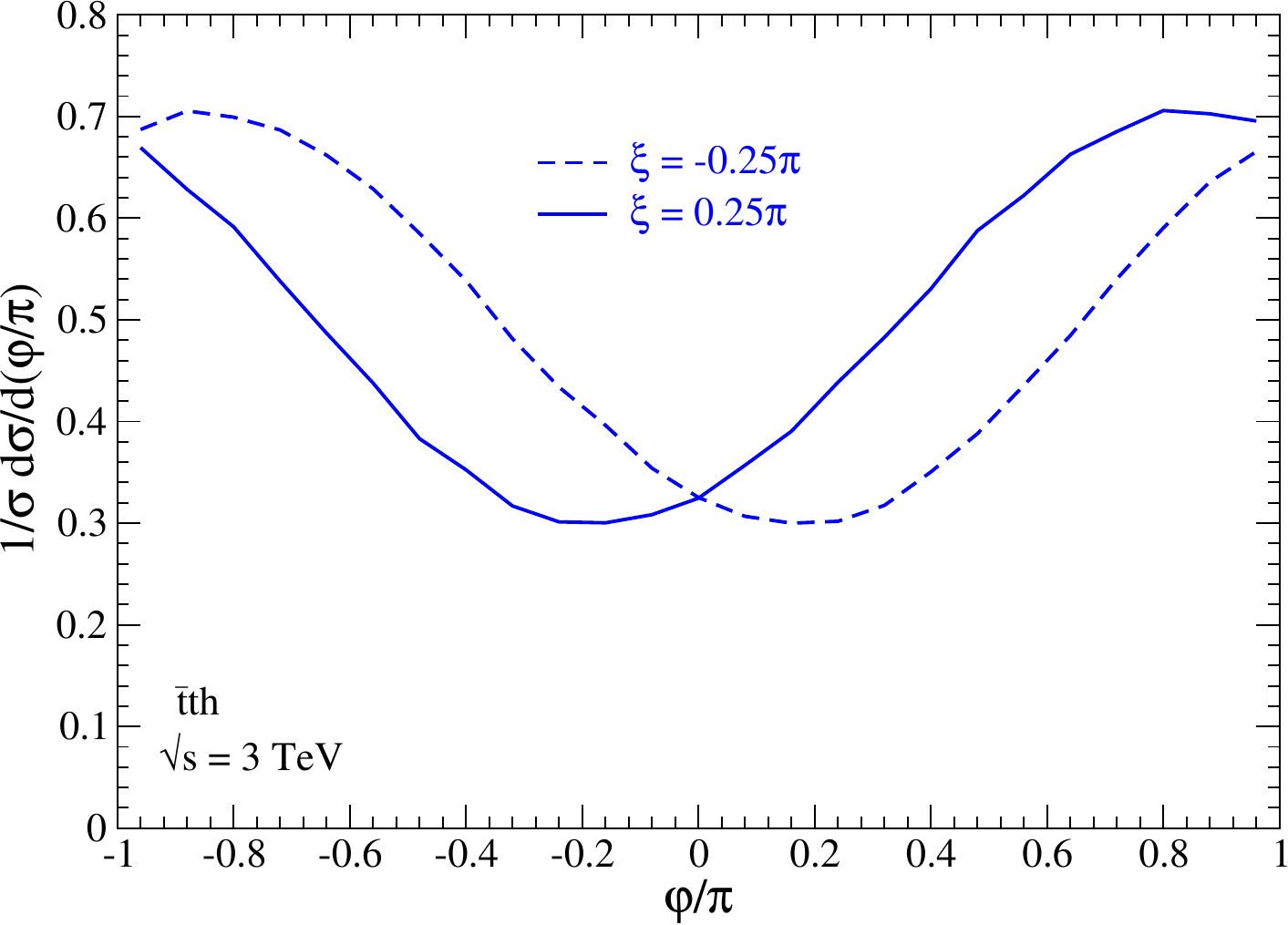}\label{fig:phit025pi}}
\hspace*{0.1cm}
\subfigure[]{\includegraphics[width=0.48\textwidth,clip]{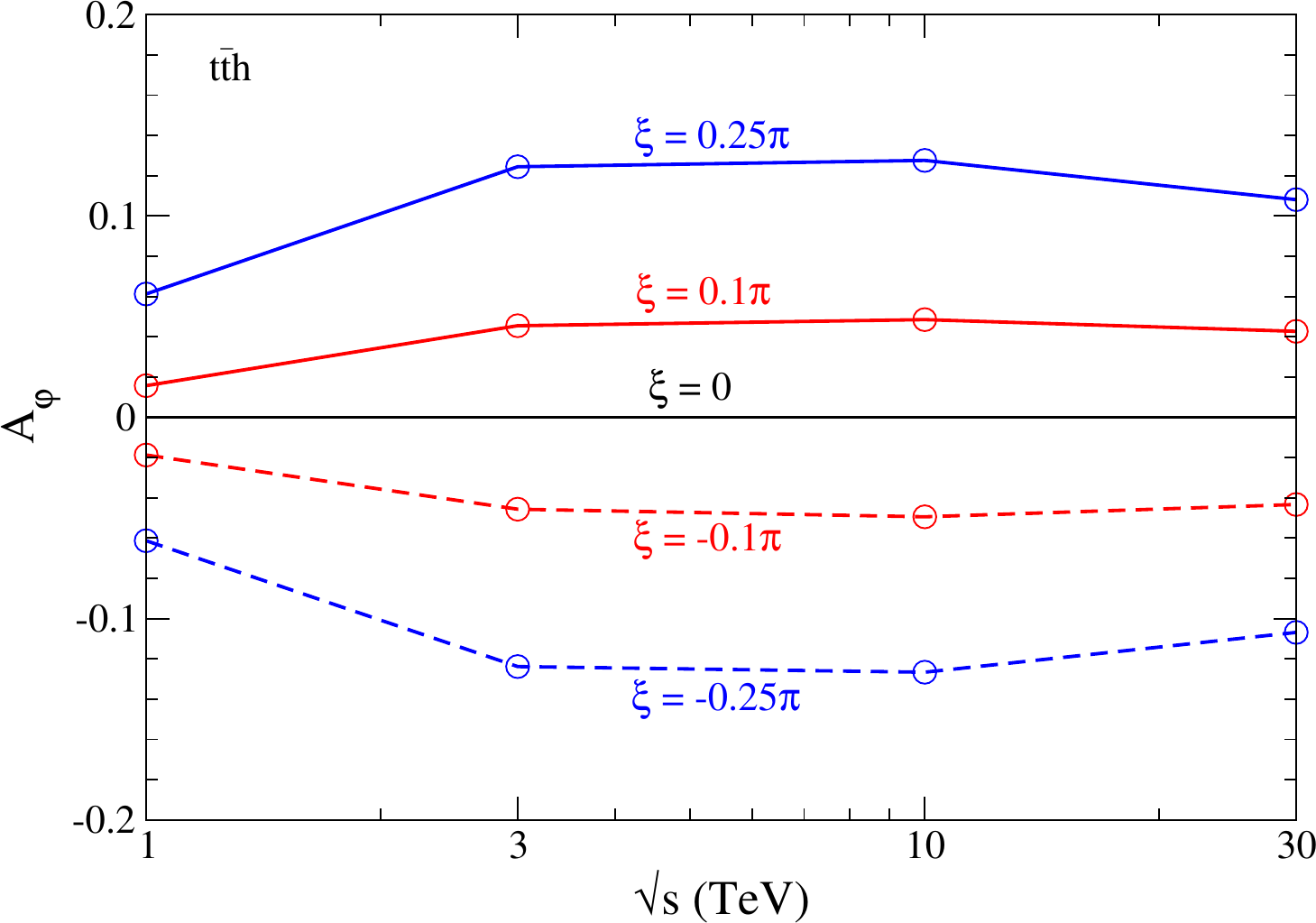}\label{fig:Aphi}}
\end{center}
\vspace*{-0.5cm}
\caption{\label{fig:phit}
 (a) Azimuthal angle ($\phi$) distribution for $t\bar{t}h$ production.  The angle is between the $t+\bar{t}$ and $h+\mu^-$ planes as defined in Eq.~(\ref{eq:azimuth}).  (b) The forward backward asymmetry of $\phi$. These are evaluated in the center of momentum frame of the $t\bar{t}h$ system.}
\end{figure}
The previous section demonstrated that at high energies the cross sections of $t\bar{t}h\nu\bar{\nu}$ and $tbh\mu\nu$ depend very strongly on the CP violating phase $\xi$.  Hence, a pure rate measurement could be very sensitive to the value of $\xi$.  However, the total cross section is CP even.  Hence, observing a deviation away from SM predictions would not demonstrate a CP violating top Yukawa.  To determine that there is indeed CP violation, we must measure observables that are sensitive to this CP violation, which we now investigate.  Additionally, as we will see in Sec.~\ref{sec:collider}, rate measurements at a muon collider with $\sqrt{s}=3$~TeV are insensitive to $\xi$.  Hence, at that energy CP violating observables may be necessary to discover direct evidence of a non-zero CP violating angle in the top-Higgs interaction.  For that reason, while the observables under consideration are general, many of the results of this section are presented for $\sqrt{s}=3$~TeV.
     
\subsubsection{$t\bar{t}h$}
We first consider the $t\bar{t}h$ process.  For this process the $t\bar{t}$ is considered as a single subsystem.  The total three-momentum of the $t\bar{t}$ subsystem is 
\begin{eqnarray}
\vec {p}_{t\bar{t}}=\vec{p}_t+\vec{p}_{\bar{t}}.
\end{eqnarray}
where $\vec{p}_t$ is the top quark momentum and $\vec{p}_{\bar{t}}$ the anti-top quark momentum.  The $t\bar{t}$ subsystem defines one plane.  Another plane can be created between the Higgs and initial state muon.  A CP odd observable can be created by considering the angle between these planes:
\begin{eqnarray}
\phi={\rm sign}\left[\vec{p}_{t\bar{t}}\cdot\left(\vec{p}_{\mu^-}\times \vec{p}_t\right)\right]\arccos\left[\frac{\vec{p}_{h\mu^-}\times \vec{p}_{\mu^-}}{|\vec{p}_{h\mu^-}\times \vec{p}_{\mu^-}|}\cdot\frac{\vec{p}_{t\bar{t}}\times\vec{p}_{t}}{|\vec{p}_{t\bar{t}}\times\vec{p}_{t}|}\right],\label{eq:azimuth}
\end{eqnarray}
where $\vec{p}_h$ is the Higgs momentum, $\vec{p}_{\mu^-}$ is the initial state $\mu^-$ momentum, and $\vec{p}_{h\mu^-}=\vec{p}_h+\vec{p}_{\mu^-}$ \cite{Atwood:1996wu,Bar-Shalom:1995quw,Gunion:1989we,Hagiwara:2017ban}.  In Fig.~\ref{fig:phit025pi} we show the  distribution of the angle $\phi$ for $\xi=\pm 0.25\pi$ and $\sqrt{s}=3$~TeV in the $t\bar{t}h$ center of momentum frame.  The distribution of $\xi=0.25\pi$ peaks at positive $\phi$ while $\xi=-0.25\pi$ peaks at negative $\phi$.  That is, there is a clear separation between $\xi=0.25\pi$ and $\xi=-0.25\pi$ and this observable is sensitive to the sign of the CP violating phase.  

Based on this observation, we define an asymmetry parameter between the regions of $\phi>0$ and $\phi<0$:
\begin{eqnarray}
A_{\phi}=\frac{\sigma(\phi>0)-\sigma(\phi<0)}{\sigma(\phi>0)+\sigma(\phi<0)}. 
\end{eqnarray}
In Fig.~\ref{fig:Aphi} we show the results of this asymmetry for $\xi=\pm0.1\pi,\pm0.25\pi$ and at energies $\sqrt{s}=1,\,3,\,10,$ and $30$~TeV.  Again, there is a clear separation with negative CP violating angles having negative asymmetry and positive CP violating angles having positive asymmetry.  Additionally, this symmetry increases between $\sqrt{s}=1$ and $3$~TeV, and slightly decreases again at $30$~TeV. To find a rough estimate on uncertainties, the benchmark points in Table~\ref{tab:lumi} are used and the statistical uncertainty on the asymmetry is estimated as $\sim 1/\sqrt{N}$. This assumes that the number of forward and backward events are roughly equal. We could expect the uncertainty of the asymmetry to be of the order $\sim 0.05$ at $1-10$ TeV and  $\sim 0.1$ at 30 TeV. 

\subsubsection{$tbh\mu\nu$ and $t\bar{t}h\nu\bar{\nu}$}

\begin{figure}[tb]
\begin{center}
{\includegraphics[width=0.55\textwidth,clip]{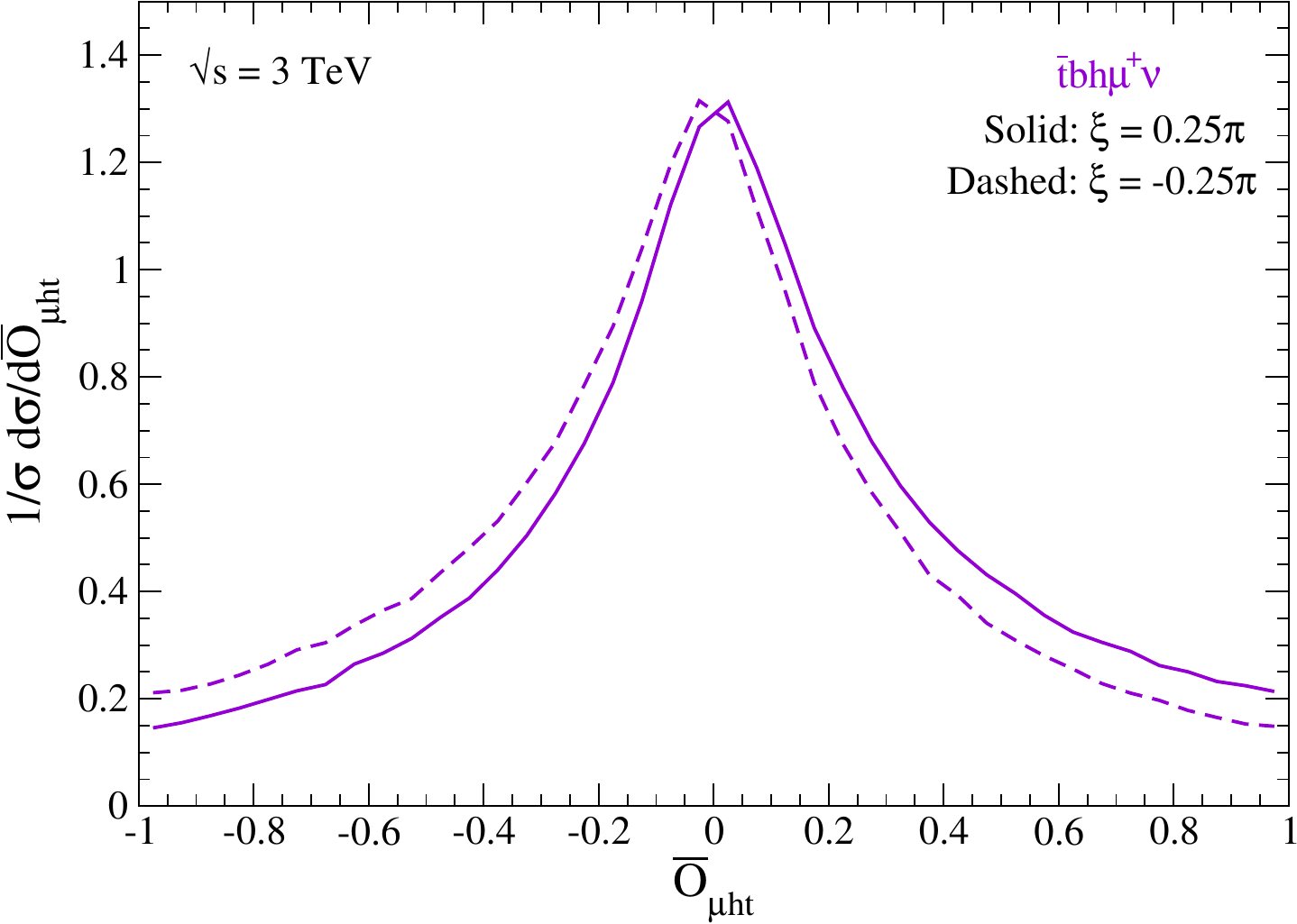}}
\end{center}
\caption{\label{fig:Omht_tbhmv}  
Differential distribution of ${\overline{\cal O}}_{\mu ht}$, defined in Eq.~(\ref{eq:Obarmuht}), for $\bar{t}bh\mu^+\nu$ at $\sqrt{s}=3$ TeV for CP violating angle (solid) $\xi=0.25\pi$ and (dashed) $\xi=-0.25\pi$.  This is evaluated in the collider center of momentum system.}
\end{figure}

\begin{figure}[tb]
\begin{center}
\subfigure[]{\includegraphics[width=0.47\textwidth,clip]{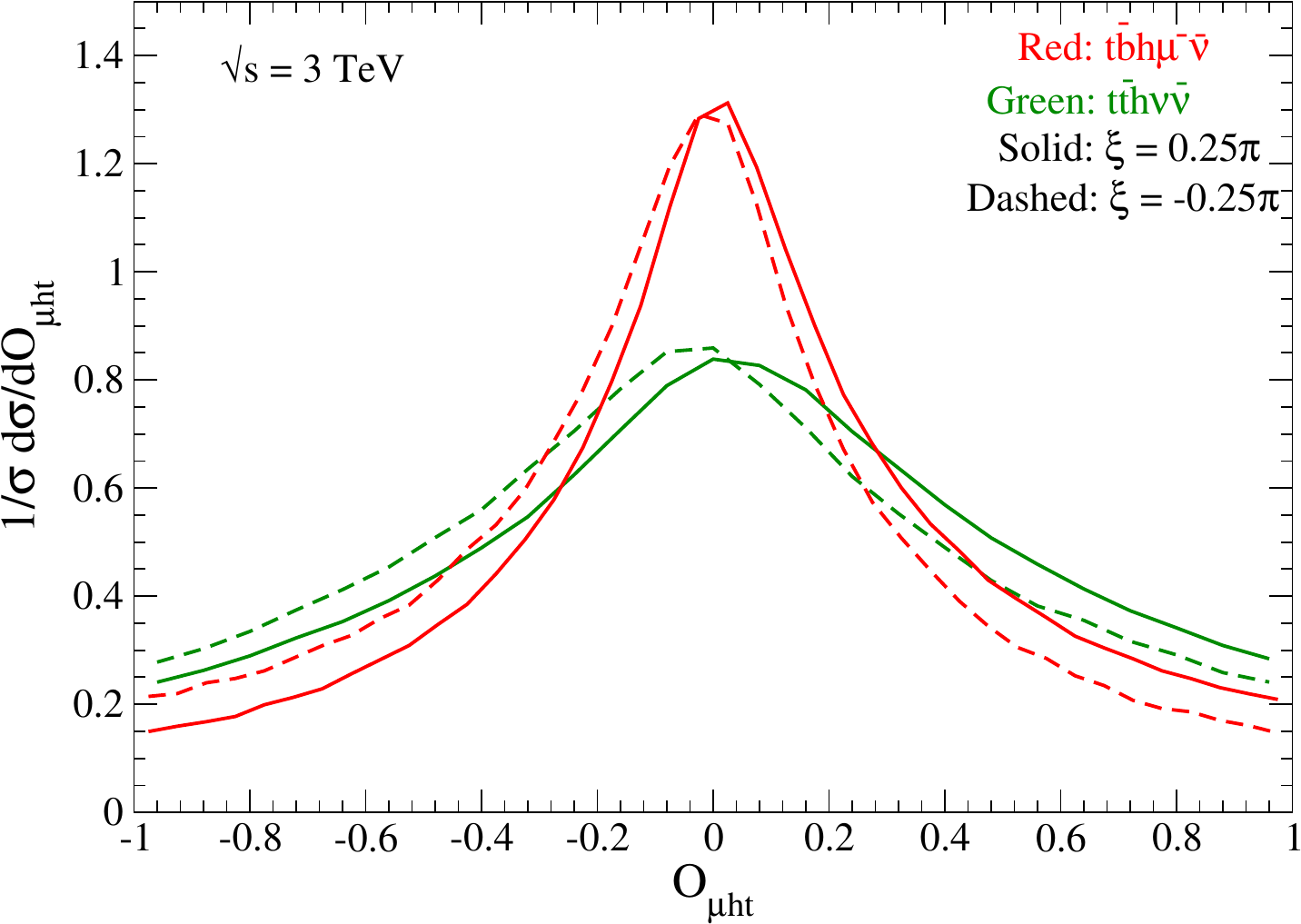}\label{fig:Omht3TeV}} \hspace*{0.1cm}
\subfigure[]{\includegraphics[width=0.49\textwidth,clip]{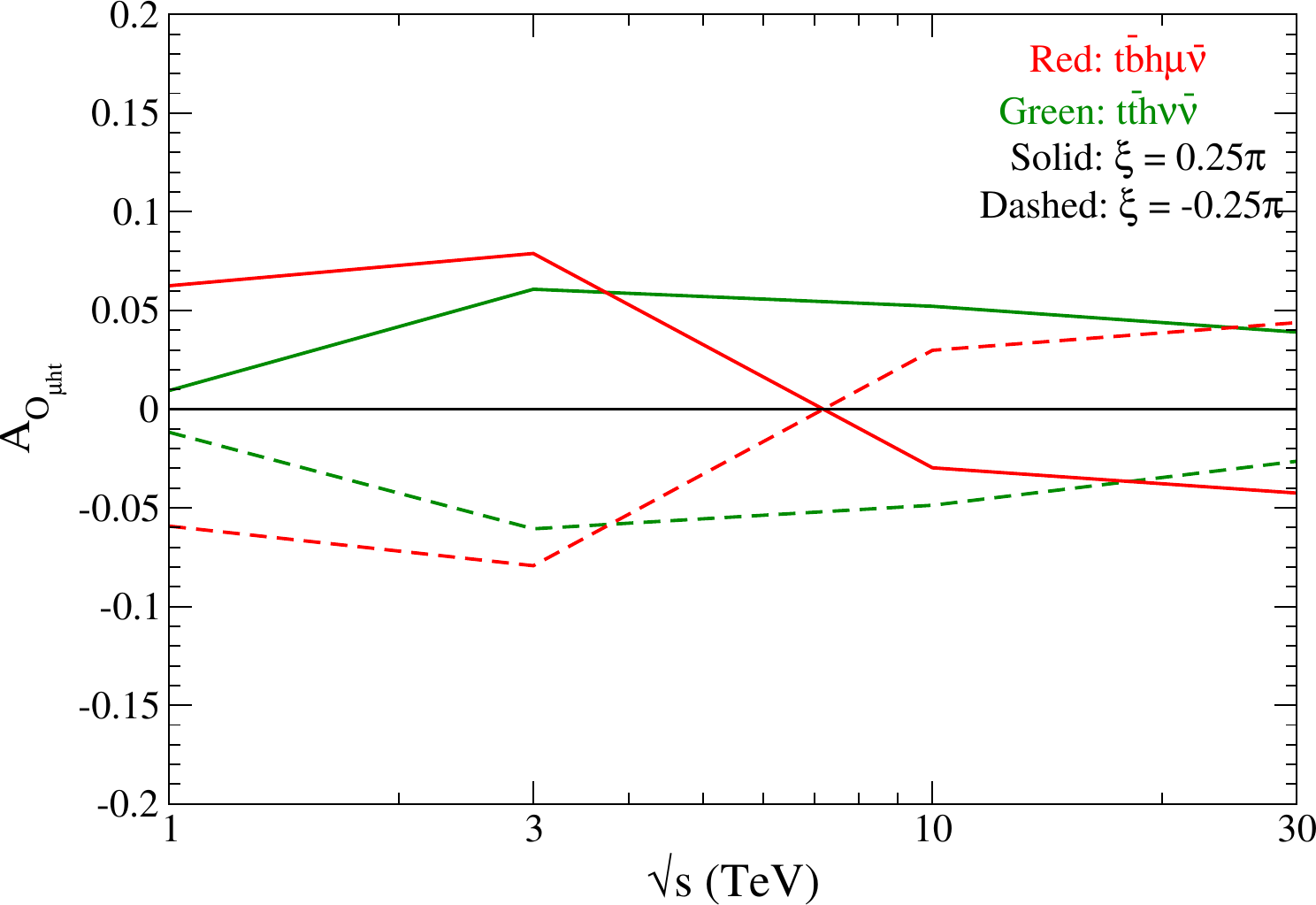}}
\end{center}
\vspace*{-0.5cm}
\caption{\label{fig:AOmht_tbhmvtthvv} (a) Differential distribution of ${\cal O}_{\mu ht}$, defined in Eq.~(\ref{eq:Omuht}), at $\sqrt{s}=3$~TeV and (b) the asymmetry parameter in Eq.~(\ref{eq:AOmuht}) as a function of $\sqrt{s}$.  These are shown for (red) $t\bar{b}h\mu^-\bar{\nu}$ and (green) $t\bar{t}h\nu\bar{\nu}$ with CP violating angle (solid) $\xi=+0.25\pi$ and (dashed) $\xi=-0.25\pi$.   These are evaluated in the collider center of momentum system.}
\end{figure}

We now turn to investigating observables sensitive to CP violation in $t\bar{t}h\nu\bar{\nu}$ and $tbh\mu\nu$ production.    
First, we consider $tbh\mu\nu$.  This process consists of $t\bar{b}h\mu^-\bar{\nu}$ and its CP conjugate $\bar{t}bh\mu^+\nu$.  We construct triple products between an initial state muon and final state products, which are odd under parity transformations~\cite{Gunion:1989we,Atwood:1996wu,Bar-Shalom:1995quw,Gunion:1996xu}.  For $t\bar{b}h\mu^-\bar{\nu}$ we use the initial state muon and final state top quark and Higgs boson:
\begin{eqnarray}
{\cal O}_{\mu ht}\equiv\frac{(\vec{p}_{\mu^-}\times\vec{p}_h)\cdot\vec{p}_t}{|\vec{p}_{\mu^-}\times\vec{p}_h||\vec{p}_t|},\label{eq:Omuht}
\end{eqnarray}
and for the conjugate process $\bar{t}bh\mu^+\nu$ we use the initial state $\mu^+$ and  final state $\bar{t}$ and Higgs boson
\begin{eqnarray}
\overline{{\cal O}}_{\mu ht}\equiv\displaystyle\frac{(\vec{p}_{\mu^+}\times\vec{p}_h)\cdot\vec{p}_{\bar{t}}}{|\vec{p}_{\mu^+}\times\vec{p}_h||\vec{p}_{\bar{t}}|}.\label{eq:Obarmuht}
\end{eqnarray}
We show the differential distributions of $\overline{\cal O}_{\mu ht}$ in Fig.~\ref{fig:Omht_tbhmv} and ${\cal O}_{\mu ht}$ in Fig.~\ref{fig:Omht3TeV} for $\sqrt{s}=3$ and $\xi=\pm0.25\pi$. The observables are evaluated in the lab frame.  There is a clear separation for positive and negative CP violating angle $\xi$, with positive (negative) $\xi$ preferring positive (negative) $\mathcal{O}_{\mu ht}$ and $\overline{\mathcal{O}}_{\mu ht}$.  Hence, these observables are sensitive to CP violation.  Since these observables are parity odd but even under charge conjugation, the distributions for the charge conjugated processes $t\bar{b}h\mu^-\overline{\nu}$ and $\bar{t}bh\mu^+\nu$ are the same.

For the $t\bar{t}h\nu\bar{\nu}$ process, we can also construct the observable $\mathcal{O}_{\mu h t}$, whose differential distribution is shown in Fig.~\ref{fig:AOmht_tbhmvtthvv}(a).  Similarly to $t\bar{b}h\mu^-\bar{\nu}$, positive $\xi$ peak at positive $\mathcal{O}_{\mu ht}$ while negative $\xi$ peak at negative $\mathcal{O}_{\mu ht}$.  However, at $\sqrt{s}=3$~TeV, the distribution is flatter for $t\bar{t}h\nu\bar{\nu}$ than it is for $t\bar{b}h\mu^-\bar{\nu}$.

Based on these observations, we define an asymmetry parameter
\begin{eqnarray}
A_{\mathcal{O}_{\mu h t}}=\frac{\sigma(\mathcal{O}_{\mu h t}>0)-\sigma(\mathcal{O}_{\mu h t}<0)}{\sigma(\mathcal{O}_{\mu h t}>0)+\sigma(\mathcal{O}_{\mu h t}<0)},\label{eq:AOmuht}
\end{eqnarray}
where 
\begin{eqnarray}
\sigma(\mathcal{O}_{\mu h t}>0)=\int_0^\infty d\mathcal{O}_{\mu h t}\frac{d\sigma}{d\mathcal{O}_{\mu h t}},
\end{eqnarray}
and similarly for $\sigma(\mathcal{O}_{\mu h t}<0)$.  The results of this asymmetry for $\sqrt{s}=1,\,3,\,10,$ and $30$~TeV are shown in Fig.~\ref{fig:AOmht_tbhmvtthvv}(b) for both $\xi=\pm 0.25\pi$ and both $t\bar{b}h\mu^-\bar{\nu}$ and $t\bar{t}h\nu\bar{\nu}$. There is a clear separation between different signs of the CP violating angle, indicating that this asymmetry is indeed sensitive to CP violation in the top Yukawa.  Interestingly, at $\sqrt{s}=1$ an $3$~TeV for $t\bar{b}\mu^-\bar{\nu}$, the CP violating angle $\xi$ and the asymmetry $A_{\mathcal{O}_{\mu ht}}$ have the same sign, while at $\sqrt{s}=10$ and $30$~TeV they have opposite sign.  For $t\bar{t}h\nu\bar{\nu}$, the sign of the asymmetry and CP violating angle are the same for all energies.

Based on the expected number of events in Table~\ref{tab:lumi}, at $\sqrt{s}=1$~TeV we could expect $\sim 0.1-0.3$ uncertainty on a measurement of $A_{\mathcal{O}_{\mu ht}}$.  Hence, a measurement of the asymmetry $A_{\phi}$ in the $tth$ system may be more sensitive to CP violation.  This is unsurprising, since $t\bar{t}h$ is the dominant production mode at $\sqrt{s}=1$~TeV.  At $\sqrt{s}=3$~TeV, the expected number of events between processes is more comparable, and the uncertainties on $A_{\mathcal{O}_{\mu h t}}$ and $A_{\phi}$ are expected to be similar: $\sim 0.05-0.1$.  At $\sqrt{s}=10$ and $30$ TeV, the $tbh\mu\nu$ and $t\bar{t}h\nu\bar{\nu}$ are dominant.  Ignoring acceptances and branching ratios, the statistical uncertainty on $A_{\mathcal{O}_{\mu h t}}$ could be expected to be percent or subpercent level, much more accurate than $A_{\phi}$.   Hence, measuring these processes at different energies could provide complementary information on the CP violating angles.  

The above estimates of uncertainties are rough and only give an order of magnitude estimation. Additionally, they do not take into account any realistic cuts needed to separate signal from background.  As shown in Ref.~\cite{Forslund:2022xjq} and as we will show below, after a collider analysis the uncertainty on a measurement of the cross section of $t\bar{t}h$, $tbh\mu\nu$, and $t\bar{t}h\nu\bar{\nu}$ is expected to be as large as $\mathcal{O}(50\%)$ at collider energies $\gtrsim 10$~TeV.  This indicates that in a realistic scenario, the uncertainty of the asymmetry parameters could be expected to be $\mathcal{O}(1)$. These are ballpark estimates to give an idea of the size of the expected statistical uncertainties. A dedicated collider analysis of the CP violating observables would be needed to determine the expected uncertainty to higher accuracy. 

%%%%%%%%%%%%%%%%%%%%%%%%%%%%%%%%%%%%%%%%%%%%%%%%%%%%%%
\section{Collider Analysis and Sensitivity Projections}
\label{sec:collider}
%%%%%%%%%%%%%%%%%%%%%%%%%%%%%%%%%%%%%%%%%%%%%%%%%%%%%%
We now present our collider analysis of the $t\bar{t}h$, $t\bar{t}h\nu\bar{\nu}$, and $tbh\mu\nu$ signals at a muon collider.  To maximize rates we consider $h\rightarrow b\bar{b}$.  In order to balance the need for rates and a clean signal, we study the semi-leptonic decay of the tops $t\rightarrow 2\,j+b$ and $t\rightarrow b\ell\nu$. Hence, our signal consists of two $b$s, two $\bar{b}$s, two jets, a lepton $\ell$, and missing transverse energy $\slashed{E}_T$ from the neutrinos:
\begin{eqnarray}
\mu^+\mu^-\rightarrow t\bar{t}h/t\bar{t}h\nu\bar{\nu}/tbh\mu\nu \rightarrow 2b+2\bar{b}+2j+\ell^\pm+\slashed{E}_T,
\end{eqnarray}
where $\ell=e,\mu$.  
Since all signal processes result in the same final state, we consider them simultaneously.  Signal and background events are simulated in \texttt{MadGraph5\_aMC@NLO}~\cite{Alwall:2014hca}, where we use new {\tt HELAS}~\cite{Hagiwara:2020tbx,Chen:2022gxv} routines to more efficiently simulate the signal.  The signal model is implemented using \texttt{FeynRules}~\cite{Christensen:2008py,Alloul:2013bka}. 

The irreducible SM background processes are
\begin{eqnarray}
\mu^-\mu^+&\to& 
gb\bar{b}/t\bar{t}g\to t\bar{t}b\bar{b},
\nonumber\\
\mu^-\mu^+&\to&t\bar{t}b\bar{b}~({\rm EW})
\nonumber\\
 \mu^-\mu^+&\to&
 t\bar{t}b\bar{b}\nu\bar{\nu}.
\end{eqnarray}
In the first line, a gluon is radiated from a $b\bar{b}$ ($t\bar{t}$) pair and then splits into $t\bar{t}$ ($b\bar{b}$). The $t\bar{t}b\bar{b}~({\rm EW})$ background is the inclusive production of $t\bar{t}b\bar{b}$ through EW processes, with no QCD contributions. 
Finally, $t\bar{t}b\bar{b}\nu\bar{\nu}$ is inclusive of all such processes.  For the $t\bar{t}b\bar{b}~({\rm EW})$ and $t\bar{t}b\bar{b}\nu\bar{\nu}$ the underlying $s$-channel, $t$-channel, and/or VBF processes are not easily disentangled and cannot be due to the necessity of gauge invariance.    Other backgrounds such as $2b2\bar{b}jj\ell^\pm+\slashed{E}_T$, $t \bar t (g\to q\bar q)$, or $t \bar t (g\to gg)$ are expected to be negligible, with requiring four $b$-tagged jets. We assume a nominal $b$-tagging efficiency of $\epsilon_b =0.9$ in our study with a percent level rate of light quark and gluon jets being mistagged as $b$-jets.~\footnote{Studies of $b$-tagging at a future muon collider suggest efficiencies of $\epsilon_b=0.5-0.6$~\cite{Bartosik:2020xwr}.  However, future projections for $b$-tagging at the HL-LHC suggest a more optimistic $b$-tagging efficiency up to $\varepsilon_b=0.85$.~\cite{CERN-LHCC-2017-005}. Hence, for our benchmarks we use an optimistic scenario.}

It should be noted that the $t\bar{t}b\bar{b}~({\rm EW})$ and $t\bar{t}b\bar{b}\nu\bar{\nu}$ background include Feynman diagrams that depend on the top Yukawa couplings and include signal diagrams.  To insure that the estimates of backgrounds do not include the signals in a gauge invariant way, we generate the backgrounds with the bottom quark Yukawa set to zero.  Additionally, these backgrounds in principle will depend on the CP structure of the top Yukawa.  The EW $t\bar{t}b\bar{b}$ background is insensitive to $\xi$.  However, at $10$ and $30$ TeV the $t\bar{t}b\bar{b}\nu\bar{\nu}$ background could increase upwards by a factor of five between $\xi=0$ and $\xi=\pm \pi/2$, and by a factor of 10 between $\xi=0$ and $\xi=\pm\pi$. This could be an indirect probe of the CP structure of the top Yukawa.  However, in this paper we are interested in direct probes.  Additionally, the importance of the background dependence on $\xi$ would depend significantly on if an analysis is performed using simulated or data driven estimates of the background.  All backgrounds that are used for our estimates are calculated for $\xi=0$.

For a more realistic analysis, we simulate the detector effects by performing a Gaussian smearing of jet energies with energy resolution 
\begin{eqnarray}
\frac{\Delta E_{b,j}}{E_{b,j}}=10\%.\label{eq:smear}
\end{eqnarray}
 For a multi-TeV lepton collider, the expected jet energy resolution can reach $3.5\%$ to $4\%$ especially for high energy jets~\cite{Linssen:2012hp,CLICdp:2018vnx,Zarnecki:2020ics}. We adopt $10\%$ as a more conservative study.  Our results show that even with a pessimistic energy resolution we can effectively distinguish signal and background.  All results presented in this section are after jet energy smearing.

For the acceptance cuts for jets and leptons,
we require them to be isolated from each other and have minimum transverse momentum, $p_T$,~\cite{CLICdp:2018vnx,Leogrande:2019qbe}:%
\begin{eqnarray}
\Delta R_{mn} >0.4,\quad p_T^{\ell,b,j}>30~{\rm GeV},\label{eq:acc}
\end{eqnarray}
where $\Delta R_{mn}=\sqrt{(\Delta\phi_{mn})^2+(\Delta\eta_{mn})^2}$ is the angular distance between particles $m$ and $n$ with $\Delta\phi_{mn}$ being the difference in azimuthal angles and $\Delta\eta_{mn}$ the rapidities.  Additionally, since our signal consists of neutrinos which appear as missing energy, we require a minimum missing transverse energy
\begin{eqnarray}
\slashed{E}_T >50~{\rm GeV}.\label{eq:met}
\end{eqnarray}

Muon colliders have significant beam induced backgrounds.  These backgrounds arise due to the finite lifetime of the muon.  Hence, they decay while in the beam producing a significant amount of charged particles along the beam direction.  To shield the detector from the beam induced backgrounds, detector proposals include shielding nozzles \cite{AlAli:2021let,Black:2022cth,MuonCollider:2022glg,MuonCollider:2022ded,MuonCollider:2022nsa,DiBenedetto:2018cpy}.
These are expected to be within $10^\circ$ of the beamline \cite{Accettura:2023ked,Liu:2023yrb,AlAli:2021let}.  We therefore apply maximum rapidity cuts on all reconstructed %charged 
particles as well:
\begin{eqnarray}
|\eta_{\ell,b,j}|<\,2.5.\label{eq:rap}
\end{eqnarray}

%%%%%%%%%%%%%%%%%%%%%%%%%%%%%%%%%%%%%%%%%%%%%%%%%%%%%%%%%%
\begin{table}[t!]
\centering
\resizebox{\linewidth}{!}{%
\begin{tabular}{c|c||c||c|c|c||ccc}\hline
$\xi$&$\sqrt{s}$ &Cuts&  \multicolumn{3}{c||}{Signal [fb]} & \multicolumn{3}{c}{Background [fb]} \\ \hline
[rad]&[TeV] & &$t\bar{t}h$   &$tbh\mu\nu$ & $t\bar{t}h\nu\bar{\nu}$ &\multirow{1}{*}{$gb\bar{b}/t\bar{t}g$}&$t\bar{t}b\bar{b}$ EW&$t\bar{t}b\bar{b}\nu\bar{\nu}$\\\hline
\multirow{6}{*}{0} &\multirow{2}{*}{1} & \multirow{1}{*}{
Acceptance} & 0.032 & 0.017 & $1.3\cdot10^{-4}$ & $6.3\cdot10^{-3}$ & 0.019 & $1.5\cdot10^{-5}$  \\\
&& \multirow{1}{*}{
$+m_h^{\rm recon}$}{} & 0.032 & 0.016 & $1.3\cdot10^{-4}$ & $3.7\cdot10^{-3}$& $8.6\cdot10^{-3}$ & $1.0\cdot10^{-5}$ \\\cline{2-9}
% 3 TeV
&\multirow{2}{*}{3} & \multirow{1}{*}{
Acceptance} & $1.3\cdot10^{-3}$ & $1.2\cdot10^{-3}$ & $2.8\cdot10^{-4}$ & $8.9\cdot10^{-4}$ & $2.6\cdot10^{-3}$ & $1.2\cdot10^{-4}$\\
&& \multirow{1}{*}{
$+m_h^{\rm recon}$}{} & $1.3\cdot10^{-3}$ & $9.4\cdot10^{-4}$ & $2.8\cdot10^{-4}$ & $1.9\cdot10^{-4}$ & $3.7\cdot10^{-4}$ & $5.5\cdot10^{-5}$ \\\cline{2-9}
% 10 TeV
&\multirow{2}{*}{10} &
Acceptance& $6.1\cdot10^{-7}$ & $1.6\cdot10^{-4}$ &$ 8.8\cdot10^{-4}$ & $2.3\cdot10^{-5}$ & $3.9\cdot10^{-5}$ & $4.2\cdot10^{-4}$ \\\
&& \multirow{1}{*}{
$+m_h^{\rm recon}$} & $5.3\cdot10^{-7}$ & $6.2\cdot10^{-5}$ & $ 8.7\cdot10^{-4}$ & $5.3\cdot10^{-7}$ & $8.2\cdot10^{-7}$ & $1.4\cdot10^{-4}$\\ \cline{2-9}
% 30 TeV
&\multirow{2}{*}{30} &
Acceptance & - & $2.2\cdot10^{-5}$ & $1.4\cdot10^{-3}$ & $2.7\cdot10^{-6}$ & $7.3\cdot10^{-7}$ & $5.2\cdot10^{-4}$\\\
&& \multirow{1}{*}{
$+m_h^{\rm recon}$} & - & $6.7\cdot10^{-6}$ & $1.4\cdot10^{-3}$ & $3.4\cdot10^{-8}$ & - & $1.6\cdot10^{-4}$\\
\hline
\hline
\multirow{6}{*}{0.1$\pi$} &\multirow{2}{*}{1} & \multirow{1}{*}{
Acceptance} & 0.031 & 0.015 & $1.3\cdot10^{-4}$ & $6.3\cdot10^{-3}$ & 0.019 & $1.5\cdot10^{-5}$ \\\
&& \multirow{1}{*}{
$+m_h^{\rm recon}$}{} & 0.031 & 0.015 & $1.3\cdot10^{-4}$ & $3.7\cdot10^{-3}$ & $8.6\cdot10^{-3}$ &$1.0\cdot10^{-5}$ \\\cline{2-9}
% 3 TeV
&\multirow{2}{*}{3} & \multirow{1}{*}{
Acceptance} & $1.4\cdot10^{-3}$ & $1.2\cdot10^{-3}$ & $6.0\cdot10^{-4}$ & $8.9\cdot10^{-4}$ & $2.6\cdot10^{-3}$ & $1.2\cdot10^{-4}$\\
&& \multirow{1}{*}{
$+m_h^{\rm recon}$}{} & $1.4\cdot10^{-3}$ & $9.0\cdot10^{-4}$ & $6.0\cdot10^{-4}$
& $1.9\cdot10^{-4}$ & $3.7\cdot10^{-4}$ & $5.5\cdot10^{-5}$ \\\cline{2-9}
% 10 TeV
&\multirow{2}{*}{10} &
Acceptance & $4.7\cdot10^{-7}$ & $1.3\cdot10^{-4}$ & $ 2.9\cdot10^{-3}$ & $2.3\cdot10^{-5}$ & $3.9\cdot10^{-5}$ & $4.2\cdot10^{-4}$\\\
&& \multirow{1}{*}{
$+m_h^{\rm recon}$} & $3.9\cdot10^{-7}$ & $5.8\cdot10^{-5}$ & $2.9\cdot10^{-3}$& $5.3\cdot10^{-7}$ & $8.2\cdot10^{-7}$ & $1.4\cdot10^{-4}$\\ \cline{2-9}
% 30 TeV
&\multirow{2}{*}{30} &
Acceptance & - & $1.7\cdot10^{-5}$ & $4.4\cdot10^{-3}$ & $2.7\cdot10^{-6}$ & $7.3\cdot10^{-7}$ & $5.2\cdot10^{-4}$\\\
&& \multirow{1}{*}{
$+m_h^{\rm recon}$} & - & $5.8\cdot10^{-6}$ & $4.4\cdot10^{-3}$ & $3.4\cdot10^{-8}$ & - & $1.6\cdot10^{-4}$\\
\hline
\hline
\multirow{6}{*}{0.2$\pi$} &\multirow{2}{*}{1} & \multirow{1}{*}{
Acceptance} & 0.025 & 0.012 & $1.1\cdot10^{-4}$& $6.3\cdot10^{-3}$ & 0.019 & $1.5\cdot10^{-5}$\\\
&& \multirow{1}{*}{
$+m_h^{\rm recon}$}{} & 0.025 & 0.012 & $1.1\cdot10^{-4}$ & $3.7\cdot10^{-3}$ & $8.6\cdot10^{-3}$ & $1.0\cdot10^{-5}$\\\cline{2-9}
% 3 TeV
&\multirow{2}{*}{3} & \multirow{1}{*}{
Acceptance} & $1.2\cdot10^{-3}$ & $1.1\cdot10^{-3}$ & $1.4\cdot10^{-3}$ & $8.9\cdot10^{-4}$ & $2.6\cdot10^{-3}$ & $1.2\cdot10^{-4}$\\
&& \multirow{1}{*}{
$+m_h^{\rm recon}$}{} & $1.2\cdot10^{-3}$& $9.1\cdot10^{-4}$ & $1.4\cdot10^{-3}$
& $1.9\cdot10^{-4}$ & $3.7\cdot10^{-4}$ & $5.5\cdot10^{-5}$ \\\cline{2-9}
% 10 TeV
&\multirow{2}{*}{10} &
Acceptance& $3.7\cdot10^{-7}$ & $1.9\cdot10^{-4}$ & $7.8\cdot10^{-3}$ & $2.3\cdot10^{-5}$ & $3.9\cdot10^{-5}$ & $4.2\cdot10^{-4}$\\\
&& \multirow{1}{*}{
$+m_h^{\rm recon}$} & $3.7\cdot10^{-7}$ & $7.8\cdot10^{-5}$ & $7.7\cdot10^{-3}$ & $5.3\cdot10^{-7}$ & $8.2\cdot10^{-7}$ & $1.4\cdot10^{-4}$\\ \cline{2-9}
% 30 TeV
&\multirow{2}{*}{30} & 
Acceptance & - & $1.8\cdot10^{-5}$ & 0.012 & $2.7\cdot10^{-6}$ & $7.3\cdot10^{-7}$ & $5.2\cdot10^{-4}$\\\
&& \multirow{1}{*}{
$+m_h^{\rm recon}$} & - & $3.7\cdot10^{-6}$ & 0.012 & $3.4\cdot10^{-8}$ & - & $1.6\cdot10^{-4}$ \\
\hline
\hline
\end{tabular}
}
\caption{\label{tab:sig} Cut flow for signal and background processes in the SM, $\xi=0.1\pi$ and $\xi=0.2\pi$. Jet energy smearing in Eq.~(\ref{eq:smear}) is applied.   Acceptance cuts consist of Eqs.~(\ref{eq:acc},~\ref{eq:met},~\ref{eq:rap}). The cut on the Higgs reconstructed mass $m_h^{\rm recon}$ are those in Eq.~(\ref{eq:mbb}). Dashes indicate negligible cross sections. We assume a nominal $b$-tagging efficiency of $\epsilon_b = 0.9$ in our study but it is not applied to cross sections in this table.
} 
\end{table}

%%%%%%%%%%%%%%%%%%%%%%%%%%%%%%%%%%%%%%%%%%%%%%%%%%%%%%%%%%

\begin{figure}[tb]
\begin{center}
\subfigure[]{\includegraphics[width=0.47\textwidth,clip]{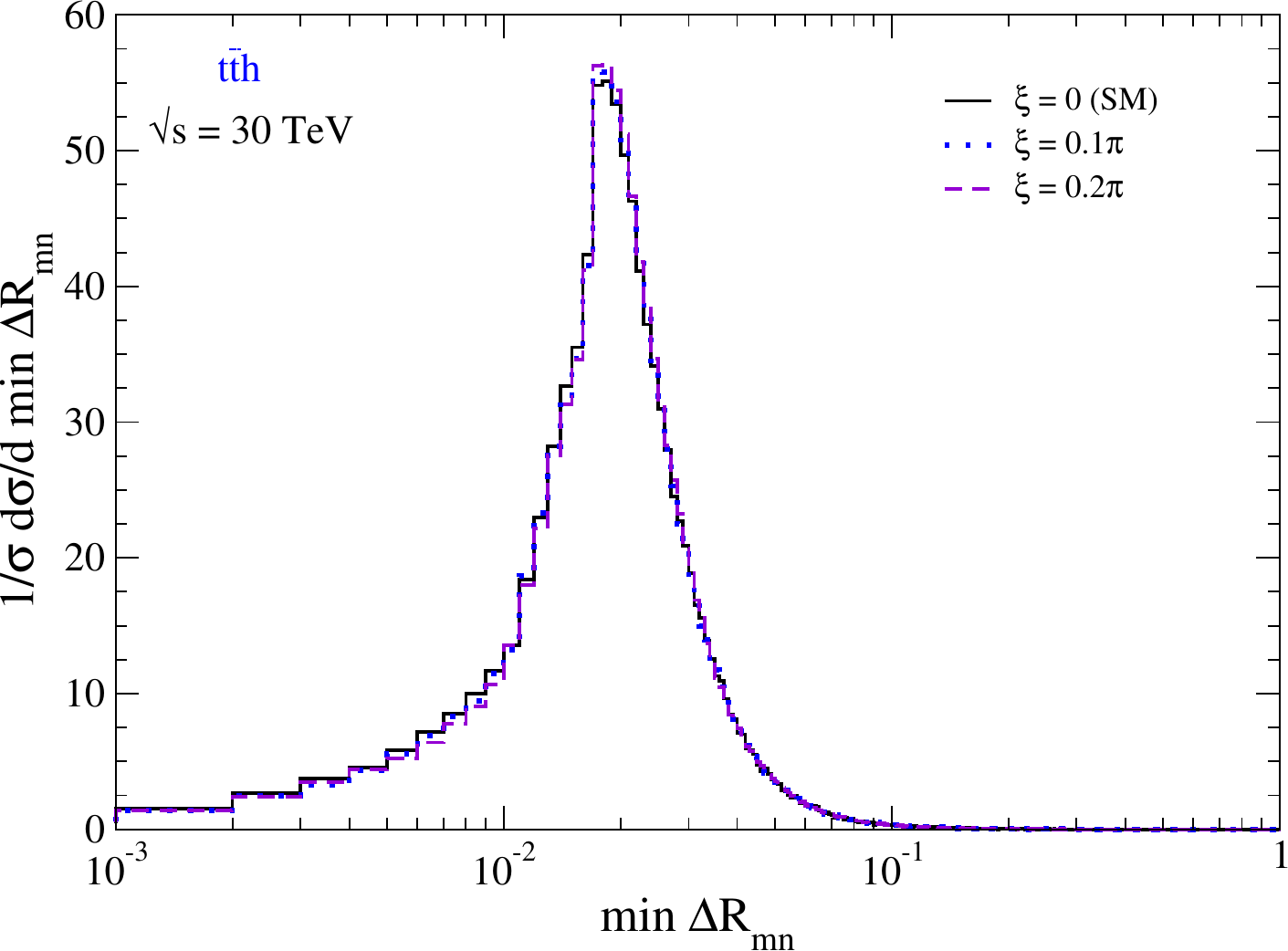}} \hspace*{0.1cm}
\subfigure[\label{fig:etal}]{\includegraphics[width=0.485\textwidth,clip]{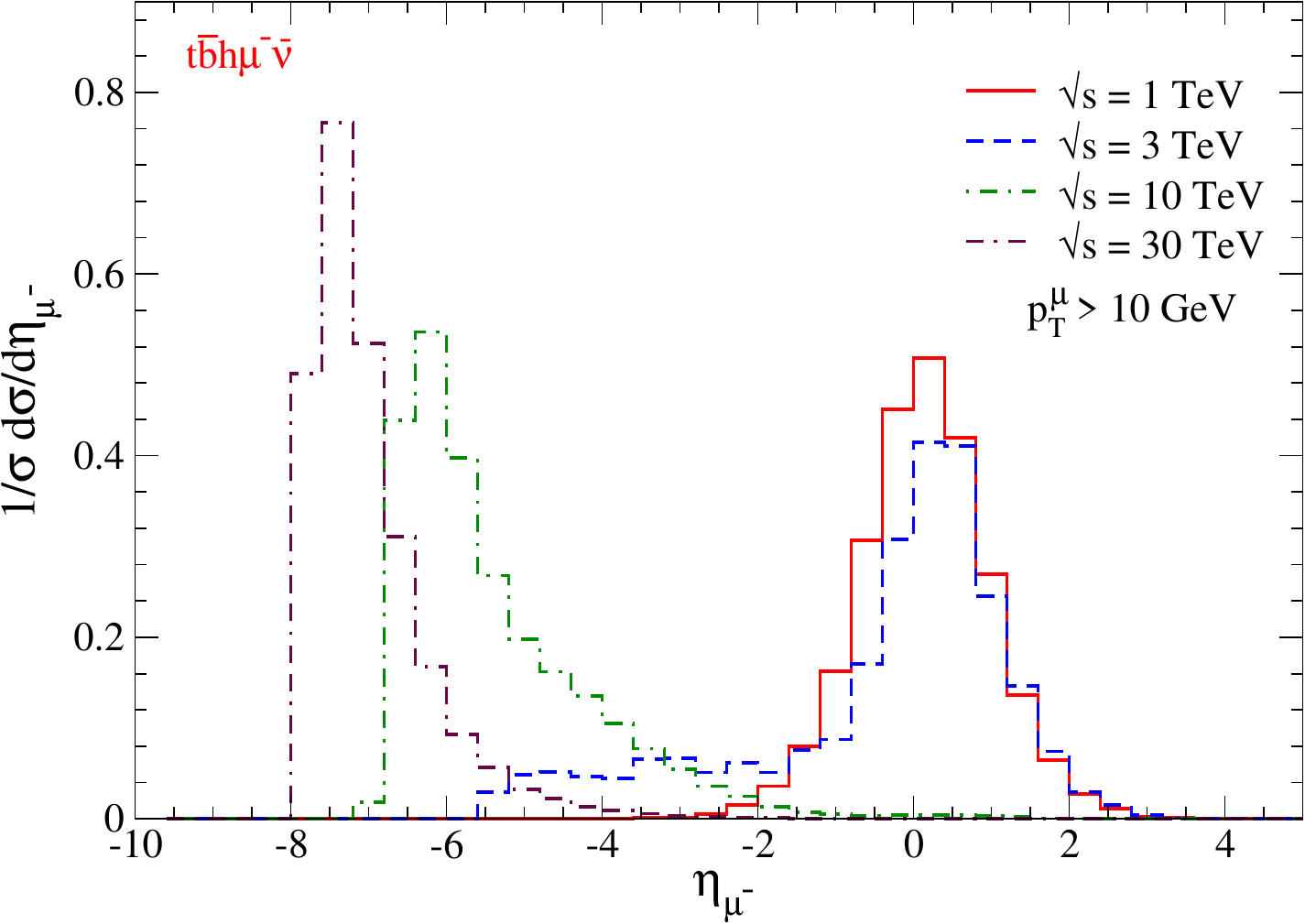}\label{fig:tbhmv_etal}}
\end{center}
\vspace*{-0.5cm}
\caption{\label{fig:dis4table} (a) Distribution of the minimal separation of $\Delta R_{mn}$ between all particles for $tth$ final state at 30 TeV.  The black solid line is for $\xi=0$, blue dotted for $\xi=0.1\pi$, and purple dashed for $\xi=0.2\pi$.  (b) Normalized rapidity distributions of final state $\mu^-$ from $t\bar{b}h\mu^-\bar{\nu}$ for $\xi=0$ at energies of (red solid) $1$~TeV, (blue dashed) $3$~TeV, (green dot-dashed) $10$~TeV, and (purple dot-dash-dashed) $30$~TeV with $p_T^\mu>10$~GeV.}
\end{figure}

In Table~\ref{tab:sig} we provide signal and background cross sections for muon collider energies of $\sqrt{s}=1,\,3,\,10,$ and $30$~TeV and CP violating angles $\xi=0,\,0.1\,\pi,$ and $0.2\,\pi$.   These cross sections are generated from 50,000 events at the generator level.\footnote{$tbh\mu\nu$ begins with fewer events at $\sqrt{s}=10,\,30$~TeV due to difficulty of generating events. However, this is a subleading signal and this does not effect our conclusion.}
After acceptance cuts and for $\xi=0$ the signal and background are comparable at $\sqrt{s}=1$, and $3$~TeV. At $\sqrt{s}=10$ and $30$ TeV, the $t\bar{t}h\nu\bar{\nu}$ signal begins to dominate the background while $tbh\mu\nu$ is relatively suppressed despite having a comparable cross section before acceptance cuts as shown in Table~\ref{tab:lumi} and Fig.~(\ref{fig:xsCPV}).  The $t\bar{t}h$ rate is greatly suppressed after acceptance cuts.  These effects can be understood with the distributions in Fig.~\ref{fig:dis4table}.  In (a) we show the minimum angular separation among all particles at $\sqrt{s}=30$~TeV for $t\bar{t}h$.  As can be clearly seen, the minimum rapidity is quite small, below the minimum required separation in Eq.~(\ref{eq:acc}).  At such high center of momentum energy, the $t\bar{t}h$ cross section is dominated by a $t\bar{t}$ topology with a Higgs radiated collinear with a parent top quark.  Since the Higgs and top are collinear, the decay products are collimated and fail the separation cuts in Eq.~(\ref{eq:acc}).

 In Fig.~\ref{fig:etal} we show the rapidity of the final state muon in $t\bar{b}h\mu^-\bar{\nu}$.   The muon is central for $\sqrt{s}=1$ and $3$ TeV, but very far backward for $\sqrt{s}=10$ and $30$~TeV.  At such high energies, this process is dominated by $t$-channel and VBF type processes.  Since the mass of vector bosons are small compared to the total energy, there are collinear $t$-channel enhancements and the VBF process is dominated by collinear emissions off the initial state $\mu^-$.  Hence, the initial and final state $\mu^-$s prefer to be in the same direction, causing the final state $\mu^-$ to be far backward in the detector\footnote{The $+z$ direction is defined as the direction of the initial state $\mu^+$, and the $-z$ the direction of the initial $\mu^-$.}.  There is a similar effect for $\bar{t}bh\mu^+\nu$ where the $\mu^+$ is very far forward at $\sqrt{s}=10$ and $30$ TeV.  Hence, at these high energies, the $tbh\mu\nu$ process largely fails the rapidity cuts in Eq.~(\ref{eq:rap}). The precise shape of these distributions depend on the cut on the final state muon transverse momentum, since that cut places a minimum on the muon rapidity.  However, the conclusions are unchanged with different transverse momentum cuts.  In these cases the signal cross section could be enhanced by tagging a forward muon.  

\subsection{Event Reconstruction and Signal/Background Separation}
\begin{figure}[tb]
\center
\includegraphics[width=0.7\textwidth,clip]{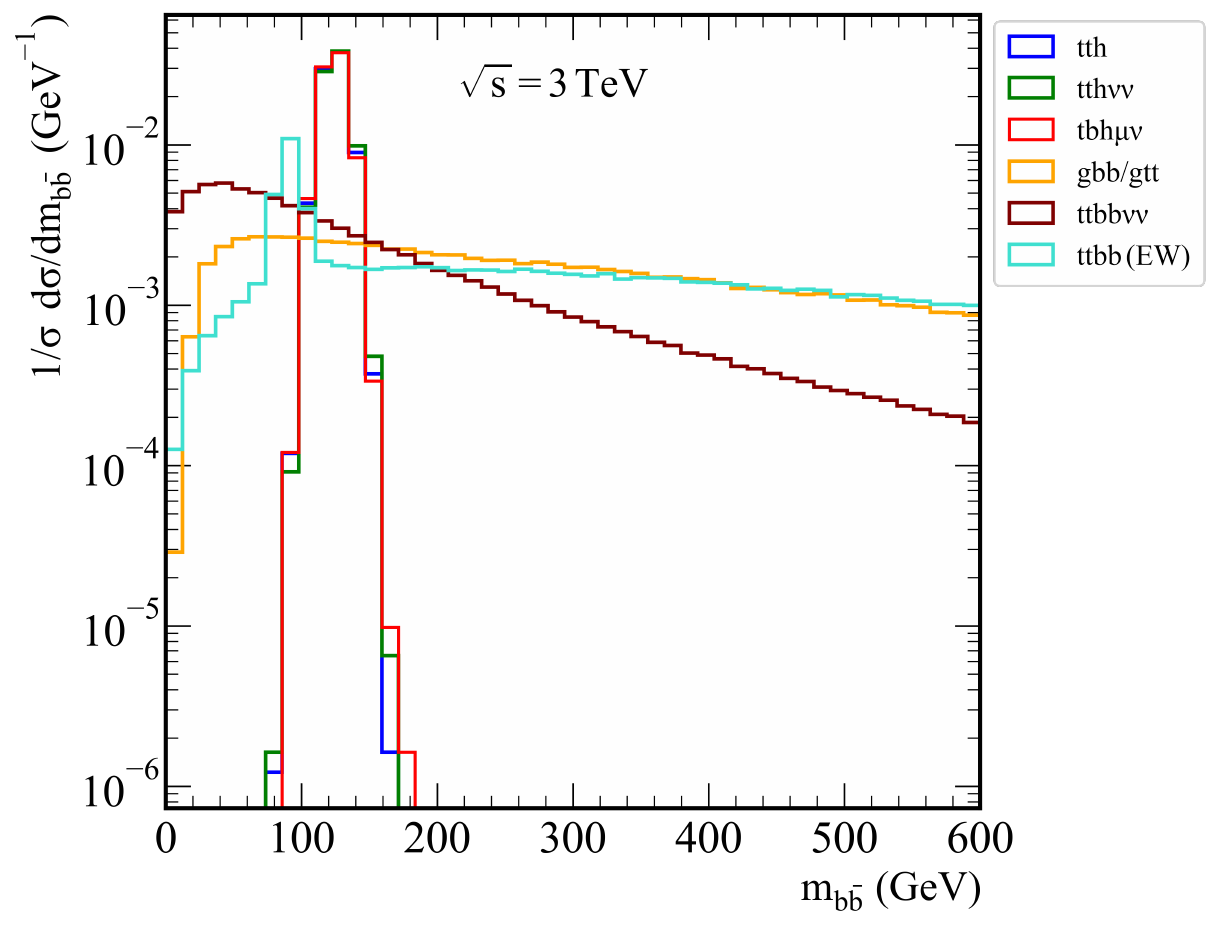}
\caption{\label{fig:mbb}Normalized $b \bar b$ invariant mass distributions for (blue, green, red) signal and (yellow, maroon, cyan) background before acceptance cuts, after jet energy smearing, and with $\xi=0$.  The signal histograms use truth level $h\rightarrow b\bar{b}$ and backgrounds bin all six possible $b$ and $\bar{b}$ combinations on an event-by-event basis. 
}
\end{figure}

The signal consists of a $b\bar{b}$ pair originating from a Higgs boson.  Hence, placing cuts on the reconstructed Higgs invariant mass can further separate signal and background.  In Fig.~\ref{fig:mbb} we show the normalized $b\bar{b}$ invariant mass distributions for $\sqrt{s}=10$~TeV.  For signal we use the truth level $h\rightarrow b\bar{b}$ decays, while for background we bin all possible bottom quark pairs.    The signals have a clear peak at $M_{b\bar{b}}=m_h=125$~GeV~\cite{Workman:2022ynf}.  The width of these peaks are dominated by the jet energy smearing due to the extreme smallness of the Higgs width $\Gamma=4.1$~MeV~\cite{LHCHiggsCrossSectionWorkingGroup:2016ypw}.  Once all combinations of $b$s  and $\bar{b}$s are considered, the $gbb/gtt$ and $t\bar{t}b\bar{b}\nu\bar{\nu}$ backgrounds produce a continuous spectrum, while the EW production of $t\bar{t}b\bar{b}$ backgrounds have $Z$-pole peaks. Even after jet energy smearing, the signal Higgs peak is separated from the background $Z$ peaks.

At the detector level, it is {\it a priori} unknown which jets originate from the top quark and Higgs. 
Hence, to use the Higgs mass for separation of signal and background, proper event reconstruction is needed.  To resolve the combinatorics in identifying the parent particles of the jets, we find the combination of jets that minimize the following $\chi^2$:  
\begin{eqnarray}
\chi^2 &= 
    \min\limits_{\vec p_\nu}&\left[\frac{(m_{b_1\ell\nu}-m_{t})^2}{\sigma_t^2} +\frac{(m_{\ell\nu}-m_W)^2}{\sigma_W^2} + \frac{(m_{b_2jj}-m_t)^2}{\sigma_t^2}\right.\nonumber\\
&&\left.
 + \frac{(|\vec p_{T,b_1\ell\nu}|-|\vec p_{T,b_2 jj}|)^2}{\sigma_{p_{T,t}}^2}
+\frac{(m_{b_3b_4}-m_H)^2}{\sigma_H^2} \right],\label{eq:chisq}
\end{eqnarray}
where $\sigma_t=5$~ GeV, $\sigma_W=5$~GeV, $\sigma_{p_{T,t}}=100$~GeV, and $\sigma_H=5$~GeV. 
The $\chi^2$ is constructed to find the combinations of particles that reconstruct the $W$ masses, top quark masses, and Higgs mass most accurately.  That is, determine the decay products of each particle. Since the base topology is $t\bar{t}$ production with a radiated Higgs, so we expect the $t$ and $\bar{t}$ to have similar transverse momenta.  However, this is very rough so the relative uncertainty on the $|\overrightarrow{p}_T|$ term is significantly larger than the mass reconstruction terms.  Note that in the case of $t\bar t h \nu \bar \nu$ production, there are two additional neutrinos from VBF processes, adding to the missing transverse momentum and therefore making the full reconstruction of the final state more challenging. However, Eq.~(\ref{eq:chisq}) works well for our purposes to disentangle the combinatorics of the problem. Also, while it is possible that a jet from a $W$-decay is misidentified as a $b$-jet, we expect the mistagging rate of light quark jets to be percent level. Since we require exactly four $b$-tagged jets, the effects of the misidentification of a jet from a $W$-decay is negligible. The above minimization procedure of $\overrightarrow{p}_\nu$ aims to find approximate three momentum of the neutrino which originates from the leptonic top quark decay without the missing transverse momentum constraint (see Refs. \cite{Huang:2022rne,Kim:2019wns,Kim:2018cxf} for similar reconstruction method using $\chi^2$), and recent review \cite{Franceschini:2022vck} on general methods for kinematic reconstruction).

Once the $b$-jets originating from the Higgs are identified, we impose a cut on reconstructed Higgs invariant mass $m_h^{\rm recon}$ around the Higgs mass $m_h=125$~GeV:
\begin{eqnarray}
|m_h^{\rm recon}-m_h|\,<\,25~{\rm GeV}.~\label{eq:mbb}
\end{eqnarray}
As shown in Table~\ref{tab:sig}, this cut does not noticeably change the total signal cross section, but greatly decreases the background.  For $\xi=0$, at $\sqrt{s}=1$~TeV the signal to background ratio increases from 1.9 to 3.9, at $3$ TeV from $0.8$  to $4.0$ , at $10$ TeV from $2.2$ to $6.4$, and at $30$ TeV from $2.6$ to $8.8$.  That is, the reconstructed Higgs mass cut can increase the signal to background ratio by a factor of two to five. The reconstructed Higgs mass cut has a small effect on the signal for $\xi=0.1\pi$ and $\xi=0.2\pi$ as well.  Hence, it increases the signal to background ratio for all $\xi$.

%%%%%
%%%%%
%%%%%%
\subsection{Significance and sensitivity to the CP violating angle}

\begin{figure}[tb]
\begin{center}
\subfigure[\label{fig:lumi_results}]{\includegraphics[width=0.52\textwidth,clip]{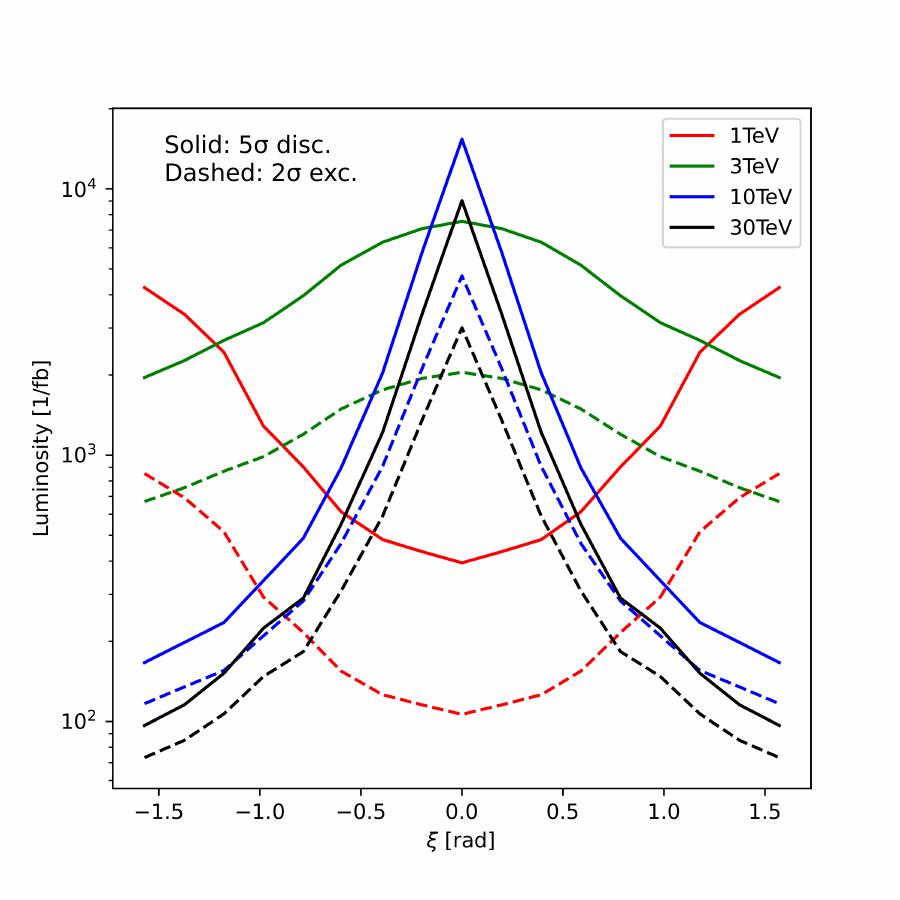}}
\hspace*{-0.9cm}
\subfigure[\label{fig:exclusion}]{\includegraphics[width=0.52\textwidth,clip]{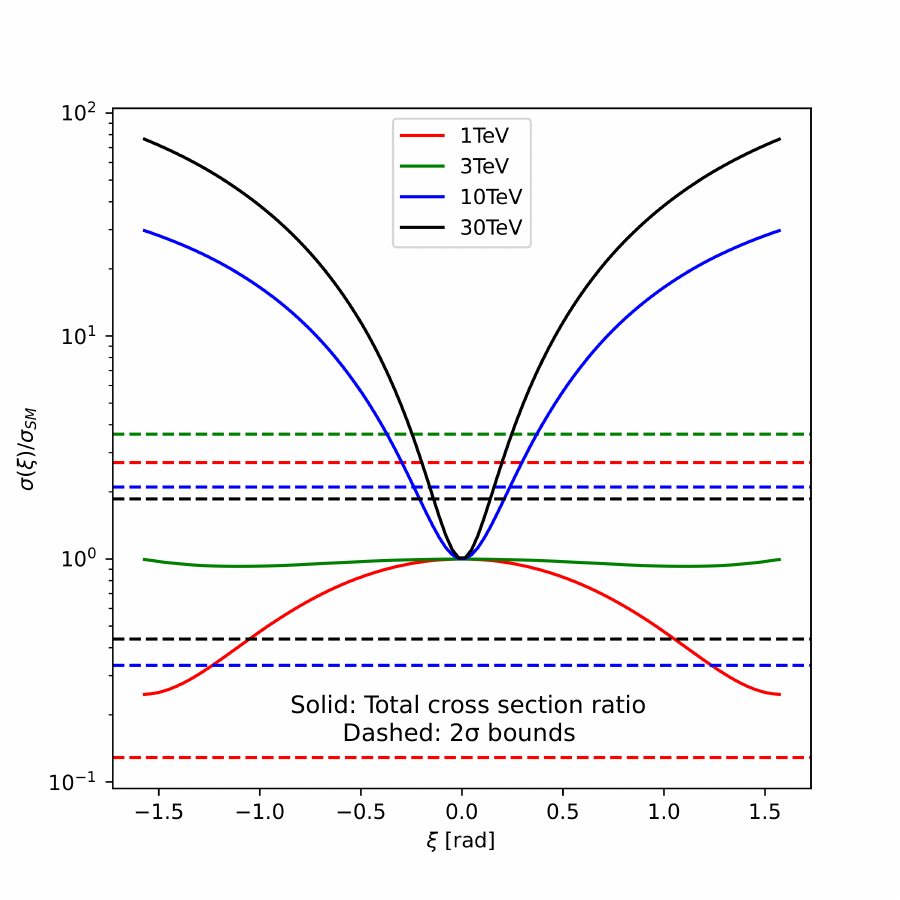}}
\end{center}
\vspace*{-0.5cm}
\caption{\label{fig:sensitivity}  (a) 
The required luminosity for 5$\sigma$ discovery (sold) and 2$\sigma$ exclusion (dashed) as a function of the CP angle.   
(b) (solid) Dependence of the ratio of the total cross section to the SM cross section on $\xi$ and (dashed) expected bounds on total cross section for the benchmark luminosities in Table~\ref{tab:sig}.  Luminosity and cross sections are shown for (red) $\sqrt{s}=1$~TeV, (green) $3$~TeV, (blue) $10$~TeV, and (black) $30$ TeV.   At $\sqrt{s}=1$ and $3$ TeV, the (solid) total cross section in (b) consists of the sum of $t\bar{t}h$, $tbh\mu\nu$, and $t\bar{t}h\nu\bar{\nu}$.  At $\sqrt{s}=10$ and $30$ TeV, the (solid) total cross section in (b) consists of $t\bar{t}h\nu\bar{\nu}$.  Results shown for $b$-tagging efficiency of $\varepsilon_b=0.9$. 
}
\end{figure}
We now present the main results of this paper.  First, we calculate the required luminosity for 2$\sigma$ exclusion and 5$\sigma$ discovery at different CP violating angles.  The discovery significance is calculated via the maximum likelihood ratio for excluding the background only hypothesis~\cite{Cowan:2010js}:
\begin{eqnarray}
\sigma_{\rm disc}=\sqrt{-2\ln\left(\frac{L(B|S+B)}{L(S+B|S+B)}\right)}\geq 5,
\end{eqnarray}
 and the exclusion significance by excluding the signal plus background hypothesis:
\begin{eqnarray}
\sigma_{\rm exc}=\sqrt{-2\ln\left(\frac{L(S+B|B)}{L(B|B)}\right)}\geq 2,
\end{eqnarray}
where $S$ is the number of signal events, $B$ is the number of background events, and $L$ is the Poisson distribution:
\begin{eqnarray}
L(x|y)=\frac{x^y}{y!}e^{-x}.
\end{eqnarray}

In Fig.~\ref{fig:lumi_results} we show the required luminosity of 5$\sigma$ discovery and 2$\sigma$ exclusion as a function of the CP violating angle $\xi$ for $\sqrt{s}=1$, $3$, $10$, and $30$~TeV.  We have included the $b$-tagging efficiencies of $\varepsilon_b=0.9$ for both signal and background.  
At $\sqrt{s}=1$~TeV, $110$~fb$^{-1}$ is needed to obtain 2$\sigma$ sensitivity to the SM rate. 
For $\xi\neq0$ the required luminosity increases. 
In this case, as shown in Fig.~\ref{fig:xsCPV}, the cross section at $\sqrt{s}=1$~TeV is maximum at $\xi=0$, and, hence, requires less luminosity to observe that $\xi\neq0$.  
For $\sqrt{s}=3$ TeV, the SM cross section is more difficult to discover or exclude than $\xi\neq 0$. Indeed, a luminosity of 
$2.1$ ab$^{-1}$ is needed for 2$\sigma$ evidence of the SM, a factor of two higher than the benchmark luminosity. This is due to the VBF diagrams beginning to make a significant contribution to the rate, where the cross sections peak at $\xi\neq0$. The effect at $10$ and $30$ TeV, where VBF typologies dominate, is even more striking. Luminosities of $4.8$ and $3$ ab$^{-1}$  are needed for  
2$\sigma$ evidence of the SM point at 10 and 30 TeV, respectively.

\begin{table}[t!]
\begin{center}
\begin{tabular}{|c||c|c|c|}\hline
\multicolumn{4}{|c|}{Projected 1$\sigma$ (2$\sigma$) bounds on $|\xi|$}\\\hline\hline
$\varepsilon_b$ & $\sqrt{s}=1$ TeV & $\sqrt{s}=10$ TeV& $\sqrt{s}=30$~TeV\\\hline
0.6 &*  &$|\xi|<15^\circ$ ($23^\circ$) & $|\xi|<8.6^\circ$ ($14^\circ$) \\
0.7 &* &  $|\xi|<12^\circ$($19^\circ$) & $|\xi|<7.2^\circ$ ($11^\circ$) \\
0.8 &$|\xi|<67^\circ$ (*) or $|\xi|>114^\circ$ (*)  & $|\xi|<10^\circ$ ($16^\circ$) & $|\xi|<6.2^\circ$ ($9.3^\circ$) \\
0.9 &$|\xi|<57^\circ$ (*) or $|\xi|>125^\circ$ (*) & $|\xi|<9.0^\circ$ ($14^\circ$) & $|\xi|<5.4^\circ$ ($8.0^\circ$)\\\hline
\end{tabular}
%}
\caption{\label{tab:XiBnds} The 1$\sigma$ projected bounds on $|\xi|$ at $\sqrt{s}=1,\,10\,$ and $30$ TeV for the benchmark luminosities in Table~\ref{tab:lumi} and a variety of $b$-tagging efficiencies $\varepsilon_b$.  The projected 2$\sigma$ bounds are in parenthesis.  At 3 TeV, there are no bounds on $\xi$ from cross section measurements.  Stars indicate no bounds on $\xi$.  
}
\end{center}
\end{table}

In Fig.~\ref{fig:exclusion}, we show the projected 2$\sigma$ constraints on the total cross section normalized to the SM value at the benchmark luminosities.  The total cross section dependence on $\xi$ is overlaid. To determine the $2\sigma$ constraints, we assume that the observations agree with the SM signal plus background.  Then we introduce a signal strength $\mu=\sigma/\sigma_{\rm SM}$ and find bounds on $\mu$ according to the likelihood:
\begin{eqnarray}
\sigma_{\rm bounds}=\sqrt{-2\log\left(\frac{L(\mu\, S_{\rm SM}+B|S_{\rm SM}+B)}{L(S_{\rm SM}+B|S_{\rm SM}+B)}\right)}\leq 2,
\end{eqnarray}
where $S_{\rm SM}$ are the expected number of SM signal events.
For the total cross section dependence on $\xi$, at $\sqrt{s}=1$ and $3$ TeV, the signal cross sections for $t\bar{t}h$, $t\bar{t}h\nu\bar{\nu}$, and $tbh\mu\nu$ are added together.  At $\sqrt{s}=10$ and $30$~TeV, as Table~\ref{tab:sig} shows, the only signal that makes a significant contribution after the analysis is $t\bar{t}h\nu\bar{\nu}$.  Hence, at these energies, we only use the $t\bar{t}h\nu\bar{\nu}$ signal to show the dependence on $\xi$.   These results consider only statistic uncertainties, but we have checked that systematic uncertainties of $5-10\%$ have little effect on our conclusions by adding them in quadrature with the statistical uncertainties.  At $2\sigma$, $\mathcal{O}(1)$ bounds can be placed on a SM-like cross section at $\sqrt{s}=1$ and $10$~TeV.  At $\sqrt{s}=3$~TeV, we find that only an upper bound on the signal cross section can be placed.  At $30$ TeV the prospects are more optimistic than other energies with projected bounds on the order of $60-80\%$.%   

In Table~\ref{tab:XiBnds} we report the projected upper bounds on $|\xi|$ for our benchmark energies and luminosities for a variety of potential $b$-tagging efficiencies of $\varepsilon_b=0.6,\,0.7,\,0.8\,$ and $0.9$~\cite{Bartosik:2020xwr,CERN-LHCC-2017-005}, which are obtained via rescaling the signal and background cross sections by $\epsilon_b^4$ corresponding to our requirement of exactly four $b$-tagged jets. As discussed previously, this requirement is to decrease the backgrounds from light quarks and gluons.  
At $1$ TeV, the cross section measurement translates to an $1\sigma$ exclusion of $125^\circ>|\xi|>57^\circ$ and no exclusion at $2\sigma$ for our benchmark $\varepsilon_b=0.9$.  At 3 TeV, the cross section measurements are insensitive to the CP violating angle and place no bounds on $\xi$.  As discussed previously, this is due to the VBF and $s$-channel processes being of similar strengths [Fig.~\ref{fig:xsCPV}].  Since the $s$-channel peaks at $\xi=0$ and the VBF processes peak at $\xi=\pm\pi$, when they are summed together the total cross is largely independent of $\xi$.  The results at $10$ and $30$ TeV are striking.  At these energies, the cross section is strongly dependent on the CP violating angle $\xi$.  Hence, even a cross section measurement with an $\mathcal{O}(50\%)$ uncertainty can exclude quite small $\xi$.  For our benchmark $b$-tagging efficiency of $\varepsilon_b=0.9$, we find that a 30 TeV (10 TeV) muon collider can exclude $|\xi|>5.4^\circ$ ($|\xi|>9.0^\circ$) at $1\sigma$ and $|\xi|>8.0^\circ$ ($|\xi|>14^\circ$) at $2\sigma$.  As the $b$-tagging efficiency decreases, the projected constraints on $\xi$ weaken.

%%%%%%%%%%%%%%%%%%%%%%%%%%%%%%%%%%%%%%%%%%%%%%%%%%%%%%
%%%%%%%%%%%%%%%%%%%%%%%%%%%%%%%%%%%%%%%%%%%%%%%%%%%%%%

%%%%%%%%%%%%%%%%%%%%%%%%%%%%%%%%%%%%%%%%%%%%%%%%%%%%%%
\section{Conclusions}
\label{sec:conc}
%%%%%%%%%%%%%%%%%%%%%%%%%%%%%%%%%%%%%%%%%%%%%%%%%%%%%%
In this paper we presented a comprehensive study of direct probes of the CP properties of the top quark Yukawa coupling at a future muon collider.  While there are many indirect probes of this coupling, direct probes are needed to provide unambiguous sensitivity to the top-Higgs coupling.  As such, we studied the $t\bar{t}h$, $t\bar{t}h\nu\bar{\nu}$, and $tbh\mu\nu$ production.  As we showed, and is well-known~\cite{Costantini:2020stv,AlAli:2021let,Han:2022edd,Aime:2022flm}, the $s$-channel process $t\bar{t}h$ is the dominant cross section at $\mathcal{O}(1~{\rm TeV})$ scale muon colliders, while the VBF type processes dominate at $\gtrsim\mathcal{O}(10~{\rm TeV})$ energies. 

The cross over between $s$-channel and VBF dominance is particularly interesting for constraining a CP violating angle. We have shown that $s$-channel and VBF processes have different dependencies on this angle.  Hence, measurements at different collider energies can provide complementary information about the CP structure of the top-Higgs interaction.  Although our total cross section measurement  including $t\bar{t}h$, $t\bar{t}h\nu\bar{\nu},$ and $tbh\mu\nu$ has $\mathcal{O}(50-100\%)$ uncertainties at the benchmark luminosities in Tab.~\ref{tab:lumi}, the constraints on the CP violating angle can be quite strong.  At $1\sigma$, rate measurements can constrain $|\xi|<57^\circ$ or $|\xi|>125^\circ$ at $\sqrt{s}=1$~TeV, $|\xi|<9^\circ$ at $\sqrt{s}=10$~TeV, and $|\xi|<5.4^\circ$ at $\sqrt{s}=30$~TeV. These constraints favorably compare to direct measurements at proposed future high energy hadron colliders~\cite{Goncalves:2021dcu,Barman:2022pip}.

As we showed in Sec.~\ref{sec:prod}, these strong constraints in VBF-style processes come from strong destructive interference between diagrams involving the top quark Yukawa and Higgs-vector boson coupling.  Hence, the additional CP violating coupling alters this destructive interference which enhances the sensitivity to $\xi$. At $\sqrt{s}=3$~TeV, the rate measurement does not meaningfully constrain the CP violating angle.  At this energy, the $s$-channel and VBF style processes have comparable cross sections. Since these processes exhibit different dependencies on $\xi$, the addition of all contributing signal rates flattens out the dependence of the total cross section on the CP violating angle.

At $\sqrt{s}=3$~TeV, measuring observables directly sensitive to CP violation may provide better sensitivity to the CP violating angle than a rate measurement.  Even at $\sqrt{s}=1,\,10,\,$ and $30$ TeV where rate measurements are sensitive to $\xi$, the rates are still CP even.  Hence, even if a first indication of a non-zero $\xi$ may appear in the $t\bar{t}h,\,t\bar{t}h\nu\bar{\nu},$ and $tbh\mu\nu$ rates, CP violating observables are needed to verify that any excess was originating from genuine CP violation.  In Sec.~\ref{sec:prod}, we explored many different CP violating observables.  For $s$-channel $t\bar{t}h$ production, the azimuthal angle between the Higgs+initial state muon plane and $t+\bar{t}$ plane is quite sensitive to the CP violating angle.  Additionally, for the VBF-style processes $t\bar{t}h\nu\bar{\nu}$ and $tbh\mu\nu$ we explored triple products to enhance sensitivity to CP violation.  Based on these results, we showed that defining asymmetries with these observables are indeed sensitive to CP violation in the top quark Yukawa.  Never-the-less, observation of a non-zero asymmetry would definitively test the CP structure of the top quark Yukawa.

While the collider analysis assumed all signal processes to be indistinguishable, the analysis of CP violating observables separated the processes.   As shown in Tab.~\ref{tab:sig}, after cuts a different processes dominate at different energies. This property could be used  to determine the most promising CP-violating observable to use. Additionally, the $t\bar{t}h$, $tbh\mu\nu$, and $t\bar{t}h\nu\bar{\nu}$ have different properties that could be used to distinguish them.  The VBF-style processes, $t\bar{t}h\nu\bar{\nu}$  and $tbh\mu\nu$, will have more missing energy than $t\bar{t}h$.  At high energies, $tbh\mu\nu$ has a very forward or backward muon while $t\bar{t}h$ and $t\bar{t}h\nu\bar{\nu}$  do not. These properties could be used to distinguish processes and guide an analysis of observable sensitive to CP-violation.  However, a dedicated collider study of the CP violating observables would be needed to determine how well this can be accomplished.

%%%%%%%%%%%%%%%%%%%%%%%%%%%%%%%%%%%%%%%%%%%%%%%%%%%%%%

\section*{Acknowledgments}
The authors would like to thank J Kanzaki and K Mawatari for instructions on new {\tt HELAS}, and D Gon\c{c}alves and K Hagiwara for useful discussions.  IML would like to thank the University of Pittsburgh Particle Physics Astrophysics and Cosmology Center and the Aspen Center for Physics, which is supported by National Science Foundation grant PHY-2210452, for their hospitality during the completion of this manuscript. 
MEC is supported in part by the Kenneth P. Dietrich School of Arts and Science Whittington Fellowship at the University of Pittsburgh.
MEC and YZ are supported in part by the State of Kansas EPSCoR grant program.  
ZD is supported in part by College of Liberal Arts and Sciences Research Fund at the University of Kansas.
KK is supported in part by US DOE DE-SC0024407. 
IML is supported in part by DE-SC0017988.  
YJZ is supported by JSPS KAKENHI Grant No.21H01077 and 23K03403. 
The data to reproduce the plots is available upon request. 

\bibliographystyle{JHEP}
\bibliography{draft}

\providecommand{\href}[2]{#2}\begingroup\raggedright\begin{thebibliography}{10}

\bibitem{Dawson:2022zbb}
S.~Dawson et~al., \emph{{Report of the Topical Group on Higgs Physics for
  Snowmass 2021: The Case for Precision Higgs Physics}},  in \emph{{2022
  Snowmass Summer Study}}, 9, 2022
  [\href{https://arxiv.org/abs/2209.07510}{{\ttfamily 2209.07510}}].

\bibitem{deBlas:2019rxi}
J.~de~Blas et~al., \emph{{Higgs Boson Studies at Future Particle Colliders}},
  \href{https://doi.org/10.1007/JHEP01(2020)139}{\emph{JHEP} {\bfseries 01}
  (2020) 139} [\href{https://arxiv.org/abs/1905.03764}{{\ttfamily
  1905.03764}}].

\bibitem{Butler:2023eah}
J.N.~Butler et~al., \emph{{Report of the 2021 U.S. Community Study on the
  Future of Particle Physics (Snowmass 2021) Summary Chapter}},  in \emph{{2022
  Snowmass Summer Study}}, 1, 2023
  [\href{https://arxiv.org/abs/2301.06581}{{\ttfamily 2301.06581}}].

\bibitem{Narain:2022qud}
M.~Narain et~al., \emph{{The Future of US Particle Physics - The Snowmass 2021
  Energy Frontier Report}},  \href{https://arxiv.org/abs/2211.11084}{{\ttfamily
  2211.11084}}.

\bibitem{CMS:2018uxb}
{\scshape CMS} collaboration, \emph{{Observation of $\mathrm{t\overline{t}}$H
  production}},
  \href{https://doi.org/10.1103/PhysRevLett.120.231801}{\emph{Phys. Rev. Lett.}
  {\bfseries 120} (2018) 231801}
  [\href{https://arxiv.org/abs/1804.02610}{{\ttfamily 1804.02610}}].

\bibitem{ATLAS:2018mme}
{\scshape ATLAS} collaboration, \emph{{Observation of Higgs boson production in
  association with a top quark pair at the LHC with the ATLAS detector}},
  \href{https://doi.org/10.1016/j.physletb.2018.07.035}{\emph{Phys. Lett. B}
  {\bfseries 784} (2018) 173}
  [\href{https://arxiv.org/abs/1806.00425}{{\ttfamily 1806.00425}}].

\bibitem{ATL-PHYS-PUB-2022-018}
{\scshape ATLAS and CMS} collaboration, \emph{{Snowmass White Paper
  Contribution: Physics with the Phase-2 ATLAS and CMS Detectors}},  Tech. Rep.
  ATL-PHYS-PUB-2022-018, CMS PAS FTR-22-001, CERN, Geneva (2022).

\bibitem{Bernardi:2022hny}
G.~Bernardi et~al., \emph{{The Future Circular Collider: a Summary for the US
  2021 Snowmass Process}},  \href{https://arxiv.org/abs/2203.06520}{{\ttfamily
  2203.06520}}.

\bibitem{CEPCPhysicsStudyGroup:2022uwl}
{\scshape CEPC Physics Study Group} collaboration, \emph{{The Physics potential
  of the CEPC. Prepared for the US Snowmass Community Planning Exercise
  (Snowmass 2021)}},  in \emph{{2022 Snowmass Summer Study}}, 5, 2022
  [\href{https://arxiv.org/abs/2205.08553}{{\ttfamily 2205.08553}}].

\bibitem{CMS:2019jdw}
{\scshape CMS} collaboration, \emph{{Constraints on anomalous $HVV$ couplings
  from the production of Higgs bosons decaying to $\tau$ lepton pairs}},
  \href{https://doi.org/10.1103/PhysRevD.100.112002}{\emph{Phys. Rev. D}
  {\bfseries 100} (2019) 112002}
  [\href{https://arxiv.org/abs/1903.06973}{{\ttfamily 1903.06973}}].

\bibitem{ATLAS:2018hxb}
{\scshape ATLAS} collaboration, \emph{{Measurements of Higgs boson properties
  in the diphoton decay channel with 36 fb$^{-1}$ of $pp$ collision data at
  $\sqrt{s} = 13$ TeV with the ATLAS detector}},
  \href{https://doi.org/10.1103/PhysRevD.98.052005}{\emph{Phys. Rev. D}
  {\bfseries 98} (2018) 052005}
  [\href{https://arxiv.org/abs/1802.04146}{{\ttfamily 1802.04146}}].

\bibitem{Buchmuller:1985jz}
W.~Buchmuller and D.~Wyler, \emph{{Effective Lagrangian Analysis of New
  Interactions and Flavor Conservation}},
  \href{https://doi.org/10.1016/0550-3213(86)90262-2}{\emph{Nucl. Phys. B}
  {\bfseries 268} (1986) 621}.

\bibitem{Grzadkowski:2010es}
B.~Grzadkowski, M.~Iskrzynski, M.~Misiak and J.~Rosiek, \emph{{Dimension-Six
  Terms in the Standard Model Lagrangian}},
  \href{https://doi.org/10.1007/JHEP10(2010)085}{\emph{JHEP} {\bfseries 10}
  (2010) 085} [\href{https://arxiv.org/abs/1008.4884}{{\ttfamily 1008.4884}}].

\bibitem{Zhang:1994fb}
X.~Zhang, S.K.~Lee, K.~Whisnant and B.L.~Young, \emph{{Phenomenology of a
  nonstandard top quark Yukawa coupling}},
  \href{https://doi.org/10.1103/PhysRevD.50.7042}{\emph{Phys. Rev. D}
  {\bfseries 50} (1994) 7042}
  [\href{https://arxiv.org/abs/hep-ph/9407259}{{\ttfamily hep-ph/9407259}}].

\bibitem{Whisnant:1994fh}
K.~Whisnant, B.-L.~Young and X.~Zhang, \emph{{Unitarity and anomalous top quark
  Yukawa couplings}},
  \href{https://doi.org/10.1103/PhysRevD.52.3115}{\emph{Phys. Rev. D}
  {\bfseries 52} (1995) 3115}
  [\href{https://arxiv.org/abs/hep-ph/9410369}{{\ttfamily hep-ph/9410369}}].

\bibitem{Whisnant:1997qu}
K.~Whisnant, J.-M.~Yang, B.-L.~Young and X.~Zhang, \emph{{Dimension-six CP
  conserving operators of the third family quarks and their effects on collider
  observables}}, \href{https://doi.org/10.1103/PhysRevD.56.467}{\emph{Phys.
  Rev. D} {\bfseries 56} (1997) 467}
  [\href{https://arxiv.org/abs/hep-ph/9702305}{{\ttfamily hep-ph/9702305}}].

\bibitem{Yang:1997iv}
J.M.~Yang and B.-L.~Young, \emph{{Dimension-six CP violating operators of the
  third family quarks and their effects at colliders}},
  \href{https://doi.org/10.1103/PhysRevD.56.5907}{\emph{Phys. Rev. D}
  {\bfseries 56} (1997) 5907}
  [\href{https://arxiv.org/abs/hep-ph/9703463}{{\ttfamily hep-ph/9703463}}].

\bibitem{Barger:2023wbg}
V.~Barger, K.~Hagiwara and Y.-J.~Zheng, \emph{{CP-violating top-Higgs coupling
  in SMEFT}}, \href{https://doi.org/10.1016/j.physletb.2024.138547}{\emph{Phys.
  Lett. B} {\bfseries 850} (2024) 138547}
  [\href{https://arxiv.org/abs/2310.10852}{{\ttfamily 2310.10852}}].

\bibitem{Brod:2022bww}
J.~Brod, J.M.~Cornell, D.~Skodras and E.~Stamou, \emph{{Global constraints on
  Yukawa operators in the standard model effective theory}},
  \href{https://doi.org/10.1007/JHEP08(2022)294}{\emph{JHEP} {\bfseries 08}
  (2022) 294} [\href{https://arxiv.org/abs/2203.03736}{{\ttfamily
  2203.03736}}].

\bibitem{Gritsan:2022php}
A.V.~Gritsan et~al., \emph{{Snowmass White Paper: Prospects of CP-violation
  measurements with the Higgs boson at future experiments}},
  \href{https://arxiv.org/abs/2205.07715}{{\ttfamily 2205.07715}}.

\bibitem{Brod:2013cka}
J.~Brod, U.~Haisch and J.~Zupan, \emph{{Constraints on CP-violating Higgs
  couplings to the third generation}},
  \href{https://doi.org/10.1007/JHEP11(2013)180}{\emph{JHEP} {\bfseries 11}
  (2013) 180} [\href{https://arxiv.org/abs/1310.1385}{{\ttfamily 1310.1385}}].

\bibitem{Bahl:2022yrs}
H.~Bahl, E.~Fuchs, S.~Heinemeyer, J.~Katzy, M.~Menen, K.~Peters et~al.,
  \emph{{Constraining the ${\mathcal {C}}{\mathcal {P}}$ structure of
  Higgs-fermion couplings with a global LHC fit, the electron EDM and
  baryogenesis}},
  \href{https://doi.org/10.1140/epjc/s10052-022-10528-1}{\emph{Eur. Phys. J. C}
  {\bfseries 82} (2022) 604}
  [\href{https://arxiv.org/abs/2202.11753}{{\ttfamily 2202.11753}}].

\bibitem{Tait:2000sh}
T.M.P.~Tait and C.P.~Yuan, \emph{{Single top quark production as a window to
  physics beyond the standard model}},
  \href{https://doi.org/10.1103/PhysRevD.63.014018}{\emph{Phys. Rev. D}
  {\bfseries 63} (2000) 014018}
  [\href{https://arxiv.org/abs/hep-ph/0007298}{{\ttfamily hep-ph/0007298}}].

\bibitem{Maltoni:2001hu}
F.~Maltoni, K.~Paul, T.~Stelzer and S.~Willenbrock, \emph{{Associated
  production of Higgs and single top at hadron colliders}},
  \href{https://doi.org/10.1103/PhysRevD.64.094023}{\emph{Phys. Rev. D}
  {\bfseries 64} (2001) 094023}
  [\href{https://arxiv.org/abs/hep-ph/0106293}{{\ttfamily hep-ph/0106293}}].

\bibitem{Barger:2009ky}
V.~Barger, M.~McCaskey and G.~Shaughnessy, \emph{{Single top and Higgs
  associated production at the LHC}},
  \href{https://doi.org/10.1103/PhysRevD.81.034020}{\emph{Phys. Rev. D}
  {\bfseries 81} (2010) 034020}
  [\href{https://arxiv.org/abs/0911.1556}{{\ttfamily 0911.1556}}].

\bibitem{Demartin:2014fia}
F.~Demartin, F.~Maltoni, K.~Mawatari, B.~Page and M.~Zaro, \emph{{Higgs
  characterisation at NLO in QCD: CP properties of the top-quark Yukawa
  interaction}},
  \href{https://doi.org/10.1140/epjc/s10052-014-3065-2}{\emph{Eur. Phys. J. C}
  {\bfseries 74} (2014) 3065}
  [\href{https://arxiv.org/abs/1407.5089}{{\ttfamily 1407.5089}}].

\bibitem{Barger:2018tqn}
V.~Barger, K.~Hagiwara and Y.-J.~Zheng, \emph{{Probing the Higgs Yukawa
  coupling to the top quark at the LHC via single top+Higgs production}},
  \href{https://doi.org/10.1103/PhysRevD.99.031701}{\emph{Phys. Rev. D}
  {\bfseries 99} (2019) 031701}
  [\href{https://arxiv.org/abs/1807.00281}{{\ttfamily 1807.00281}}].

\bibitem{Barger:2019ccj}
V.~Barger, K.~Hagiwara and Y.-J.~Zheng, \emph{{Probing the top Yukawa coupling
  at the LHC via associated production of single top and Higgs}},
  \href{https://doi.org/10.1007/JHEP09(2020)101}{\emph{JHEP} {\bfseries 09}
  (2020) 101} [\href{https://arxiv.org/abs/1912.11795}{{\ttfamily
  1912.11795}}].

\bibitem{ATLAS:2020ior}
{\scshape ATLAS} collaboration, \emph{{$CP$ Properties of Higgs Boson
  Interactions with Top Quarks in the $t\bar{t}H$ and $tH$ Processes Using $H
  \rightarrow \gamma\gamma$ with the ATLAS Detector}},
  \href{https://doi.org/10.1103/PhysRevLett.125.061802}{\emph{Phys. Rev. Lett.}
  {\bfseries 125} (2020) 061802}
  [\href{https://arxiv.org/abs/2004.04545}{{\ttfamily 2004.04545}}].

\bibitem{ATLAS:2023cbt}
{\scshape ATLAS} collaboration, \emph{{Probing the CP nature of the
  top\textendash{}Higgs Yukawa coupling in tt\textasciimacron{}H and tH events
  with H\textrightarrow{}bb\textasciimacron{} decays using the ATLAS detector
  at the LHC}},
  \href{https://doi.org/10.1016/j.physletb.2024.138469}{\emph{Phys. Lett. B}
  {\bfseries 849} (2024) 138469}
  [\href{https://arxiv.org/abs/2303.05974}{{\ttfamily 2303.05974}}].

\bibitem{CMS:2022dbt}
{\scshape CMS} collaboration, \emph{{Search for $CP$ violation in ttH and tH
  production in multilepton channels in proton-proton collisions at $\sqrt{s}$
  = 13 TeV}}, \href{https://doi.org/10.1007/JHEP07(2023)092}{\emph{JHEP}
  {\bfseries 07} (2023) 092}
  [\href{https://arxiv.org/abs/2208.02686}{{\ttfamily 2208.02686}}].

\bibitem{CMS:2015nrd}
{\scshape CMS} collaboration, \emph{{Search for the associated production of a
  Higgs boson with a single top quark in proton-proton collisions at $
  \sqrt{s}=8 $ TeV}},
  \href{https://doi.org/10.1007/JHEP06(2016)177}{\emph{JHEP} {\bfseries 06}
  (2016) 177} [\href{https://arxiv.org/abs/1509.08159}{{\ttfamily
  1509.08159}}].

\bibitem{CMS:2018jeh}
{\scshape CMS} collaboration, \emph{{Search for associated production of a
  Higgs boson and a single top quark in proton-proton collisions at $\sqrt{s} =
  13$ TeV}}, \href{https://doi.org/10.1103/PhysRevD.99.092005}{\emph{Phys. Rev.
  D} {\bfseries 99} (2019) 092005}
  [\href{https://arxiv.org/abs/1811.09696}{{\ttfamily 1811.09696}}].

\bibitem{Ellis:2013yxa}
J.~Ellis, D.S.~Hwang, K.~Sakurai and M.~Takeuchi, \emph{{Disentangling
  Higgs-Top Couplings in Associated Production}},
  \href{https://doi.org/10.1007/JHEP04(2014)004}{\emph{JHEP} {\bfseries 04}
  (2014) 004} [\href{https://arxiv.org/abs/1312.5736}{{\ttfamily 1312.5736}}].

\bibitem{Boudjema:2015nda}
F.~Boudjema, R.M.~Godbole, D.~Guadagnoli and K.A.~Mohan, \emph{{Lab-frame
  observables for probing the top-Higgs interaction}},
  \href{https://doi.org/10.1103/PhysRevD.92.015019}{\emph{Phys. Rev. D}
  {\bfseries 92} (2015) 015019}
  [\href{https://arxiv.org/abs/1501.03157}{{\ttfamily 1501.03157}}].

\bibitem{Buckley:2015vsa}
M.R.~Buckley and D.~Goncalves, \emph{{Boosting the Direct CP Measurement of the
  Higgs-Top Coupling}},
  \href{https://doi.org/10.1103/PhysRevLett.116.091801}{\emph{Phys. Rev. Lett.}
  {\bfseries 116} (2016) 091801}
  [\href{https://arxiv.org/abs/1507.07926}{{\ttfamily 1507.07926}}].

\bibitem{Gritsan:2016hjl}
A.V.~Gritsan, R.~R\"ontsch, M.~Schulze and M.~Xiao, \emph{{Constraining
  anomalous Higgs boson couplings to the heavy flavor fermions using matrix
  element techniques}},
  \href{https://doi.org/10.1103/PhysRevD.94.055023}{\emph{Phys. Rev. D}
  {\bfseries 94} (2016) 055023}
  [\href{https://arxiv.org/abs/1606.03107}{{\ttfamily 1606.03107}}].

\bibitem{Mileo:2016mxg}
N.~Mileo, K.~Kiers, A.~Szynkman, D.~Crane and E.~Gegner, \emph{{Pseudoscalar
  top-Higgs coupling: exploration of CP-odd observables to resolve the sign
  ambiguity}}, \href{https://doi.org/10.1007/JHEP07(2016)056}{\emph{JHEP}
  {\bfseries 07} (2016) 056}
  [\href{https://arxiv.org/abs/1603.03632}{{\ttfamily 1603.03632}}].

\bibitem{AmorDosSantos:2017ayi}
S.~Amor Dos~Santos et~al., \emph{{Probing the CP nature of the Higgs coupling
  in $t{\bar t}h$ events at the LHC}},
  \href{https://doi.org/10.1103/PhysRevD.96.013004}{\emph{Phys. Rev. D}
  {\bfseries 96} (2017) 013004}
  [\href{https://arxiv.org/abs/1704.03565}{{\ttfamily 1704.03565}}].

\bibitem{Azevedo:2017qiz}
D.~Azevedo, A.~Onofre, F.~Filthaut and R.~Gon\c{c}alo, \emph{{CP tests of Higgs
  couplings in $t\bar{t}h$ semileptonic events at the LHC}},
  \href{https://doi.org/10.1103/PhysRevD.98.033004}{\emph{Phys. Rev. D}
  {\bfseries 98} (2018) 033004}
  [\href{https://arxiv.org/abs/1711.05292}{{\ttfamily 1711.05292}}].

\bibitem{Li:2017dyz}
J.~Li, Z.-g.~Si, L.~Wu and J.~Yue, \emph{{Central-edge asymmetry as a probe of
  Higgs-top coupling in $t\bar{t}h$ production at the LHC}},
  \href{https://doi.org/10.1016/j.physletb.2018.02.009}{\emph{Phys. Lett. B}
  {\bfseries 779} (2018) 72}
  [\href{https://arxiv.org/abs/1701.00224}{{\ttfamily 1701.00224}}].

\bibitem{Goncalves:2018agy}
D.~Gon\c{c}alves, K.~Kong and J.H.~Kim, \emph{{Probing the top-Higgs Yukawa CP
  structure in dileptonic $ t\overline{t}h $ with M$_{2}$-assisted
  reconstruction}}, \href{https://doi.org/10.1007/JHEP06(2018)079}{\emph{JHEP}
  {\bfseries 06} (2018) 079}
  [\href{https://arxiv.org/abs/1804.05874}{{\ttfamily 1804.05874}}].

\bibitem{Ren:2019xhp}
J.~Ren, L.~Wu and J.M.~Yang, \emph{{Unveiling CP property of top-Higgs coupling
  with graph neural networks at the LHC}},
  \href{https://doi.org/10.1016/j.physletb.2020.135198}{\emph{Phys. Lett. B}
  {\bfseries 802} (2020) 135198}
  [\href{https://arxiv.org/abs/1901.05627}{{\ttfamily 1901.05627}}].

\bibitem{Bortolato:2020zcg}
B.~Bortolato, J.F.~Kamenik, N.~Ko\v{s}nik and A.~Smolkovi\v{c},
  \emph{{Optimized probes of $CP$ -odd effects in the $t \bar t h$ process at
  hadron colliders}},
  \href{https://doi.org/10.1016/j.nuclphysb.2021.115328}{\emph{Nucl. Phys. B}
  {\bfseries 964} (2021) 115328}
  [\href{https://arxiv.org/abs/2006.13110}{{\ttfamily 2006.13110}}].

\bibitem{Cao:2020hhb}
Q.-H.~Cao, K.-P.~Xie, H.~Zhang and R.~Zhang, \emph{{A New Observable for
  Measuring CP Property of Top-Higgs Interaction}},
  \href{https://doi.org/10.1088/1674-1137/abcfac}{\emph{Chin. Phys. C}
  {\bfseries 45} (2021) 023117}
  [\href{https://arxiv.org/abs/2008.13442}{{\ttfamily 2008.13442}}].

\bibitem{Martini:2021uey}
T.~Martini, R.-Q.~Pan, M.~Schulze and M.~Xiao, \emph{{Probing the CP structure
  of the top quark Yukawa coupling: Loop sensitivity versus on-shell
  sensitivity}}, \href{https://doi.org/10.1103/PhysRevD.104.055045}{\emph{Phys.
  Rev. D} {\bfseries 104} (2021) 055045}
  [\href{https://arxiv.org/abs/2104.04277}{{\ttfamily 2104.04277}}].

\bibitem{Barman:2021yfh}
R.K.~Barman, D.~Gon\c{c}alves and F.~Kling, \emph{{Machine learning the Higgs
  boson-top quark CP phase}},
  \href{https://doi.org/10.1103/PhysRevD.105.035023}{\emph{Phys. Rev. D}
  {\bfseries 105} (2022) 035023}
  [\href{https://arxiv.org/abs/2110.07635}{{\ttfamily 2110.07635}}].

\bibitem{Barman:2022pip}
R.K.~Barman et~al., \emph{{Directly Probing the CP-structure of the Higgs-Top
  Yukawa at HL-LHC and Future Colliders}},  in \emph{{2022 Snowmass Summer
  Study}}, 3, 2022 [\href{https://arxiv.org/abs/2203.08127}{{\ttfamily
  2203.08127}}].

\bibitem{BhupalDev:2007ftb}
P.S.~Bhupal~Dev, A.~Djouadi, R.M.~Godbole, M.M.~Muhlleitner and S.D.~Rindani,
  \emph{{Determining the CP properties of the Higgs boson}},
  \href{https://doi.org/10.1103/PhysRevLett.100.051801}{\emph{Phys. Rev. Lett.}
  {\bfseries 100} (2008) 051801}
  [\href{https://arxiv.org/abs/0707.2878}{{\ttfamily 0707.2878}}].

\bibitem{Goncalves:2021dcu}
D.~Gon\c{c}alves, J.H.~Kim, K.~Kong and Y.~Wu, \emph{{Direct Higgs-top CP-phase
  measurement with $ t\overline{t}h $ at the 14 TeV LHC and 100 TeV FCC}},
  \href{https://doi.org/10.1007/JHEP01(2022)158}{\emph{JHEP} {\bfseries 01}
  (2022) 158} [\href{https://arxiv.org/abs/2108.01083}{{\ttfamily
  2108.01083}}].

\bibitem{Ackerschott:2023nax}
J.~Ackerschott, R.K.~Barman, D.~Gon\c{c}alves, T.~Heimel and T.~Plehn,
  \emph{{Returning CP-Observables to The Frames They Belong}},
  \href{https://arxiv.org/abs/2308.00027}{{\ttfamily 2308.00027}}.

\bibitem{Heimel:2023mvw}
T.~Heimel, N.~Huetsch, R.~Winterhalder, T.~Plehn and A.~Butter,
  \emph{{Precision-Machine Learning for the Matrix Element Method}},
  \href{https://arxiv.org/abs/2310.07752}{{\ttfamily 2310.07752}}.

\bibitem{Palmer:1995jy}
R.B.~Palmer et~al., \emph{{Muon colliders}},
  \href{https://doi.org/10.1063/1.50922}{\emph{AIP Conf. Proc.} {\bfseries 372}
  (1996) 3} [\href{https://arxiv.org/abs/acc-phys/9602001}{{\ttfamily
  acc-phys/9602001}}].

\bibitem{Palmer:2014nza}
R.B.~Palmer, \emph{{Muon Colliders}},
  \href{https://doi.org/10.1142/S1793626814300072}{\emph{Rev. Accel. Sci.
  Tech.} {\bfseries 7} (2014) 137}.

\bibitem{Boscolo:2018ytm}
M.~Boscolo, J.-P.~Delahaye and M.~Palmer, \emph{{The future prospects of muon
  colliders and neutrino factories}},
  \href{https://doi.org/10.1142/9789811209604_0010}{\emph{Rev. Accel. Sci.
  Tech.} {\bfseries 10} (2019) 189}
  [\href{https://arxiv.org/abs/1808.01858}{{\ttfamily 1808.01858}}].

\bibitem{Delahaye:2019omf}
J.P.~Delahaye, M.~Diemoz, K.~Long, B.~Mansouli\'e, N.~Pastrone, L.~Rivkin
  et~al., \emph{{Muon Colliders}},
  \href{https://arxiv.org/abs/1901.06150}{{\ttfamily 1901.06150}}.

\bibitem{AlAli:2021let}
H.~Al~Ali et~al., \emph{{The muon Smasher\textquoteright{}s guide}},
  \href{https://doi.org/10.1088/1361-6633/ac6678}{\emph{Rept. Prog. Phys.}
  {\bfseries 85} (2022) 084201}
  [\href{https://arxiv.org/abs/2103.14043}{{\ttfamily 2103.14043}}].

\bibitem{Franceschini:2021aqd}
R.~Franceschini and M.~Greco, \emph{{Higgs and BSM Physics at the Future Muon
  Collider}}, \href{https://doi.org/10.3390/sym13050851}{\emph{Symmetry}
  {\bfseries 13} (2021) 851}
  [\href{https://arxiv.org/abs/2104.05770}{{\ttfamily 2104.05770}}].

\bibitem{Black:2022cth}
K.M.~Black et~al., \emph{{Muon Collider Forum Report}},
  \href{https://arxiv.org/abs/2209.01318}{{\ttfamily 2209.01318}}.

\bibitem{MuonCollider:2022nsa}
{\scshape Muon Collider} collaboration, \emph{{A Muon Collider Facility for
  Physics Discovery}},  \href{https://arxiv.org/abs/2203.08033}{{\ttfamily
  2203.08033}}.

\bibitem{MuonCollider:2022xlm}
{\scshape Muon Collider} collaboration, \emph{{The physics case of a 3 TeV muon
  collider stage}},  \href{https://arxiv.org/abs/2203.07261}{{\ttfamily
  2203.07261}}.

\bibitem{Aime:2022flm}
C.~Aime et~al., \emph{{Muon Collider Physics Summary}},
  \href{https://arxiv.org/abs/2203.07256}{{\ttfamily 2203.07256}}.

\bibitem{Accettura:2023ked}
C.~Accettura et~al., \emph{{Towards a muon collider}},
  \href{https://doi.org/10.1140/epjc/s10052-023-11889-x}{\emph{Eur. Phys. J. C}
  {\bfseries 83} (2023) 864}
  [\href{https://arxiv.org/abs/2303.08533}{{\ttfamily 2303.08533}}].

\bibitem{MuonCollider:2022glg}
{\scshape Muon Collider} collaboration, \emph{{Promising Technologies and R\&D
  Directions for the Future Muon Collider Detectors}},
  \href{https://arxiv.org/abs/2203.07224}{{\ttfamily 2203.07224}}.

\bibitem{MuonCollider:2022ded}
{\scshape Muon Collider} collaboration, \emph{{Simulated Detector Performance
  at the Muon Collider}},  \href{https://arxiv.org/abs/2203.07964}{{\ttfamily
  2203.07964}}.

\bibitem{Forslund:2022xjq}
M.~Forslund and P.~Meade, \emph{{High precision higgs from high energy muon
  colliders}}, \href{https://doi.org/10.1007/JHEP08(2022)185}{\emph{JHEP}
  {\bfseries 08} (2022) 185}
  [\href{https://arxiv.org/abs/2203.09425}{{\ttfamily 2203.09425}}].

\bibitem{Costantini:2020stv}
A.~Costantini, F.~De~Lillo, F.~Maltoni, L.~Mantani, O.~Mattelaer, R.~Ruiz
  et~al., \emph{{Vector boson fusion at multi-TeV muon colliders}},
  \href{https://doi.org/10.1007/JHEP09(2020)080}{\emph{JHEP} {\bfseries 09}
  (2020) 080} [\href{https://arxiv.org/abs/2005.10289}{{\ttfamily
  2005.10289}}].

\bibitem{Han:2022edd}
T.~Han, S.~Li, S.~Su, W.~Su and Y.~Wu, \emph{{BSM Higgs Production at a Muon
  Collider}},  in \emph{{2022 Snowmass Summer Study}}, 5, 2022
  [\href{https://arxiv.org/abs/2205.11730}{{\ttfamily 2205.11730}}].

\bibitem{Chen:2022yiu}
M.~Chen and D.~Liu, \emph{{Top Yukawa coupling measurement at the muon
  collider}}, \href{https://doi.org/10.1103/PhysRevD.109.075020}{\emph{Phys.
  Rev. D} {\bfseries 109} (2024) 075020}
  [\href{https://arxiv.org/abs/2212.11067}{{\ttfamily 2212.11067}}].

\bibitem{Liu:2023yrb}
Z.~Liu, K.-F.~Lyu, I.~Mahbub and L.-T.~Wang, \emph{{Top Yukawa coupling
  determination at high energy muon collider}},
  \href{https://doi.org/10.1103/PhysRevD.109.035021}{\emph{Phys. Rev. D}
  {\bfseries 109} (2024) 035021}
  [\href{https://arxiv.org/abs/2308.06323}{{\ttfamily 2308.06323}}].

\bibitem{Alwall:2014hca}
J.~Alwall, R.~Frederix, S.~Frixione, V.~Hirschi, F.~Maltoni, O.~Mattelaer
  et~al., \emph{{The automated computation of tree-level and next-to-leading
  order differential cross sections, and their matching to parton shower
  simulations}}, \href{https://doi.org/10.1007/JHEP07(2014)079}{\emph{JHEP}
  {\bfseries 07} (2014) 079} [\href{https://arxiv.org/abs/1405.0301}{{\ttfamily
  1405.0301}}].

\bibitem{Dawson:1984gx}
S.~Dawson, \emph{{The Effective W Approximation}},
  \href{https://doi.org/10.1016/0550-3213(85)90038-0}{\emph{Nucl. Phys. B}
  {\bfseries 249} (1985) 42}.

\bibitem{Ruiz:2021tdt}
R.~Ruiz, A.~Costantini, F.~Maltoni and O.~Mattelaer, \emph{{The Effective
  Vector Boson Approximation in high-energy muon collisions}},
  \href{https://doi.org/10.1007/JHEP06(2022)114}{\emph{JHEP} {\bfseries 06}
  (2022) 114} [\href{https://arxiv.org/abs/2111.02442}{{\ttfamily
  2111.02442}}].

\bibitem{Han:2020uid}
T.~Han, Y.~Ma and K.~Xie, \emph{{High energy leptonic collisions and
  electroweak parton distribution functions}},
  \href{https://doi.org/10.1103/PhysRevD.103.L031301}{\emph{Phys. Rev. D}
  {\bfseries 103} (2021) L031301}
  [\href{https://arxiv.org/abs/2007.14300}{{\ttfamily 2007.14300}}].

\bibitem{vonWeizsacker:1934nji}
C.F.~von Weizsacker, \emph{{Radiation emitted in collisions of very fast
  electrons}}, \href{https://doi.org/10.1007/BF01333110}{\emph{Z. Phys.}
  {\bfseries 88} (1934) 612}.

\bibitem{Williams:1935dka}
E.J.~Williams, \emph{{Correlation of certain collision problems with radiation
  theory}}, {\emph{Kong. Dan. Vid. Sel. Mat. Fys. Med.} {\bfseries 13N4} (1935)
  1}.

\bibitem{Christensen:2008py}
N.D.~Christensen and C.~Duhr, \emph{{FeynRules - Feynman rules made easy}},
  \href{https://doi.org/10.1016/j.cpc.2009.02.018}{\emph{Comput. Phys. Commun.}
  {\bfseries 180} (2009) 1614}
  [\href{https://arxiv.org/abs/0806.4194}{{\ttfamily 0806.4194}}].

\bibitem{Alloul:2013bka}
A.~Alloul, N.D.~Christensen, C.~Degrande, C.~Duhr and B.~Fuks, \emph{{FeynRules
  2.0 - A complete toolbox for tree-level phenomenology}},
  \href{https://doi.org/10.1016/j.cpc.2014.04.012}{\emph{Comput. Phys. Commun.}
  {\bfseries 185} (2014) 2250}
  [\href{https://arxiv.org/abs/1310.1921}{{\ttfamily 1310.1921}}].

\bibitem{Atwood:1996wu}
D.~Atwood and A.~Soni, \emph{{CP violation in top physics at the NLC}},  in
  \emph{{28th International Conference on High-energy Physics}},
  pp.~1119--1122, 7, 1996
  [\href{https://arxiv.org/abs/hep-ph/9609418}{{\ttfamily hep-ph/9609418}}].

\bibitem{Bar-Shalom:1995quw}
S.~Bar-Shalom, D.~Atwood, G.~Eilam, R.R.~Mendel and A.~Soni, \emph{{Large tree
  level CP violation in $e^{+} e^{-} \to t \bar{t} H^0$ in the two Higgs
  doublet model}}, \href{https://doi.org/10.1103/PhysRevD.53.1162}{\emph{Phys.
  Rev. D} {\bfseries 53} (1996) 1162}
  [\href{https://arxiv.org/abs/hep-ph/9508314}{{\ttfamily hep-ph/9508314}}].

\bibitem{Gunion:1989we}
J.F.~Gunion, H.E.~Haber, G.L.~Kane and S.~Dawson, \emph{{The Higgs Hunter's
  Guide}}, vol.~80 (2000).

\bibitem{Hagiwara:2017ban}
K.~Hagiwara, H.~Yokoya and Y.-J.~Zheng, \emph{{Probing the CP properties of top
  Yukawa coupling at an $e^+e^-$ collider}},
  \href{https://doi.org/10.1007/JHEP02(2018)180}{\emph{JHEP} {\bfseries 02}
  (2018) 180} [\href{https://arxiv.org/abs/1712.09953}{{\ttfamily
  1712.09953}}].

\bibitem{Gunion:1996xu}
J.F.~Gunion and X.-G.~He, \emph{{Determining the CP nature of a neutral Higgs
  boson at the LHC}},
  \href{https://doi.org/10.1103/PhysRevLett.76.4468}{\emph{Phys. Rev. Lett.}
  {\bfseries 76} (1996) 4468}
  [\href{https://arxiv.org/abs/hep-ph/9602226}{{\ttfamily hep-ph/9602226}}].

\bibitem{Hagiwara:2020tbx}
K.~Hagiwara, J.~Kanzaki and K.~Mawatari, \emph{{QED and QCD helicity amplitudes
  in parton-shower gauge}},
  \href{https://doi.org/10.1140/epjc/s10052-020-8154-9}{\emph{Eur. Phys. J. C}
  {\bfseries 80} (2020) 584}
  [\href{https://arxiv.org/abs/2003.03003}{{\ttfamily 2003.03003}}].

\bibitem{Chen:2022gxv}
J.~Chen, K.~Hagiwara, J.~Kanzaki and K.~Mawatari, \emph{{Helicity amplitudes
  without gauge cancellation for electroweak processes}},
  \href{https://doi.org/10.1140/epjc/s10052-023-12093-7}{\emph{Eur. Phys. J. C}
  {\bfseries 83} (2023) 922}
  [\href{https://arxiv.org/abs/2203.10440}{{\ttfamily 2203.10440}}].

\bibitem{Bartosik:2020xwr}
N.~Bartosik et~al., \emph{{Detector and Physics Performance at a Muon
  Collider}},
  \href{https://doi.org/10.1088/1748-0221/15/05/P05001}{\emph{JINST} {\bfseries
  15} (2020) P05001} [\href{https://arxiv.org/abs/2001.04431}{{\ttfamily
  2001.04431}}].

\bibitem{CERN-LHCC-2017-005}
{\scshape ATLAS Collaboration} collaboration, \emph{{Technical Design Report
  for the ATLAS Inner Tracker Strip Detector}},  Tech. Rep.
  \href{https://cds.cern.ch/record/2257755}{CERN-LHCC-2017-005, ATLAS-TDR-025},
  CERN, Geneva (Apr, 2017).

\bibitem{Linssen:2012hp}
L.~Linssen, A.~Miyamoto, M.~Stanitzki and H.~Weerts, eds., \emph{{Physics and
  Detectors at CLIC: CLIC Conceptual Design Report}},
  \href{https://arxiv.org/abs/1202.5940}{{\ttfamily 1202.5940}}.

\bibitem{CLICdp:2018vnx}
{\scshape CLICdp} collaboration, \emph{{A detector for CLIC: main parameters
  and performance}},  \href{https://arxiv.org/abs/1812.07337}{{\ttfamily
  1812.07337}}.

\bibitem{Zarnecki:2020ics}
{\scshape CLICdp, ILD concept group} collaboration, \emph{{On the physics
  potential of ILC and CLIC}},
  \href{https://doi.org/10.22323/1.376.0037}{\emph{PoS} {\bfseries CORFU2019}
  (2020) 037} [\href{https://arxiv.org/abs/2004.14628}{{\ttfamily
  2004.14628}}].

\bibitem{Leogrande:2019qbe}
E.~Leogrande, P.~Roloff, U.~Schnoor and M.~Weber, \emph{{A DELPHES card for the
  CLIC detector}},  \href{https://arxiv.org/abs/1909.12728}{{\ttfamily
  1909.12728}}.

\bibitem{DiBenedetto:2018cpy}
V.~Di~Benedetto, C.~Gatto, A.~Mazzacane, N.V.~Mokhov, S.I.~Striganov and
  N.K.~Terentiev, \emph{{A Study of Muon Collider Background Rejection Criteria
  in Silicon Vertex and Tracker Detectors}},
  \href{https://doi.org/10.1088/1748-0221/13/09/P09004}{\emph{JINST} {\bfseries
  13} (2018) P09004} [\href{https://arxiv.org/abs/1807.00074}{{\ttfamily
  1807.00074}}].

\bibitem{Workman:2022ynf}
{\scshape Particle Data Group} collaboration, \emph{{Review of Particle
  Physics}}, \href{https://doi.org/10.1093/ptep/ptac097}{\emph{PTEP} {\bfseries
  2022} (2022) 083C01}.

\bibitem{LHCHiggsCrossSectionWorkingGroup:2016ypw}
{\scshape LHC Higgs Cross Section Working Group} collaboration, \emph{{Handbook
  of LHC Higgs Cross Sections: 4. Deciphering the Nature of the Higgs Sector}},
   \href{https://arxiv.org/abs/1610.07922}{{\ttfamily 1610.07922}}.

\bibitem{Huang:2022rne}
L.~Huang, S.-b.~Kang, J.H.~Kim, K.~Kong and J.S.~Pi, \emph{{Portraying double
  Higgs at the Large Hadron Collider II}},
  \href{https://doi.org/10.1007/JHEP08(2022)114}{\emph{JHEP} {\bfseries 08}
  (2022) 114} [\href{https://arxiv.org/abs/2203.11951}{{\ttfamily
  2203.11951}}].

\bibitem{Kim:2019wns}
J.H.~Kim, M.~Kim, K.~Kong, K.T.~Matchev and M.~Park, \emph{{Portraying Double
  Higgs at the Large Hadron Collider}},
  \href{https://doi.org/10.1007/JHEP09(2019)047}{\emph{JHEP} {\bfseries 09}
  (2019) 047} [\href{https://arxiv.org/abs/1904.08549}{{\ttfamily
  1904.08549}}].

\bibitem{Kim:2018cxf}
J.H.~Kim, K.~Kong, K.T.~Matchev and M.~Park, \emph{{Probing the Triple Higgs
  Self-Interaction at the Large Hadron Collider}},
  \href{https://doi.org/10.1103/PhysRevLett.122.091801}{\emph{Phys. Rev. Lett.}
  {\bfseries 122} (2019) 091801}
  [\href{https://arxiv.org/abs/1807.11498}{{\ttfamily 1807.11498}}].

\bibitem{Franceschini:2022vck}
R.~Franceschini, D.~Kim, K.~Kong, K.T.~Matchev, M.~Park and P.~Shyamsundar,
  \emph{{Kinematic variables and feature engineering for particle
  phenomenology}},
  \href{https://doi.org/10.1103/RevModPhys.95.045004}{\emph{Rev. Mod. Phys.}
  {\bfseries 95} (2023) 045004}
  [\href{https://arxiv.org/abs/2206.13431}{{\ttfamily 2206.13431}}].

\bibitem{Cowan:2010js}
G.~Cowan, K.~Cranmer, E.~Gross and O.~Vitells, \emph{{Asymptotic formulae for
  likelihood-based tests of new physics}},
  \href{https://doi.org/10.1140/epjc/s10052-011-1554-0}{\emph{Eur. Phys. J. C}
  {\bfseries 71} (2011) 1554}
  [\href{https://arxiv.org/abs/1007.1727}{{\ttfamily 1007.1727}}].

\end{thebibliography}\endgroup

\end{document}